\documentclass[reprint,twocolumn,superscriptaddress,showpacs,showkeys,amsmath,amssymb,floatfix,aps,prb]{revtex4-1}

\usepackage{graphicx}
\usepackage{dcolumn}
\usepackage{bm}
\usepackage{amsmath}
\usepackage{placeins}
\usepackage{color}
\usepackage{hyperref}
\usepackage{multirow}

\begin{document}

\title{Heterostructures of graphene and hBN: electronic, spin-orbit, and spin relaxation properties from first principles}

\author{Klaus Zollner}
	\email{klaus.zollner@physik.uni-regensburg.de}
	\affiliation{Institute for Theoretical Physics, University of Regensburg, 93040 Regensburg, Germany}	
\author{Martin Gmitra}
	\affiliation{Institute of Physics, P. J. \v{S}af\'{a}rik University in Ko\v{s}ice, 04001 Ko\v{s}ice, Slovakia}
\author{Jaroslav Fabian}
	\affiliation{Institute for Theoretical Physics, University of Regensburg, 93040 Regensburg, Germany}
\date{\today}

\begin{abstract}
We perform extensive first-principles calculations for heterostructures 
composed of monolayer graphene and hexagonal boron nitride (hBN). 
Employing a symmetry-derived minimal tight-binding model, 
we extract orbital and spin-orbit coupling (SOC) parameters 
for graphene on hBN, as well as for hBN encapsulated graphene.
Our calculations show that the parameters depend on the 
specific stacking configuration of graphene on hBN. 
We also perform an interlayer distance study for the 
different graphene/hBN stacks to find
the corresponding lowest energy distances. 
For very large interlayer distances, one can recover
the pristine graphene properties, as we find from the 
dependence of the parameters on the interlayer distance. 
Furthermore, we find that orbital and SOC parameters, especially the Rashba one, 
depend strongly on an applied transverse electric field, giving a rich playground for spin physics.
Armed with the model parameters, we employ the Dyakonov-Perel formalism 
to calculate the spin relaxation in graphene/hBN heterostructures.
We find spin lifetimes in the nanosecond range, in agreement with recent measurements. 
The spin relaxation anisotropy, being the ratio of out-of-plane to in-plane spin lifetimes, 
is found to be giant close to the charge neutrality point, decreasing with increasing doping, 
and being highly tunable by an external transverse electric field. 
This is in contrast to bilayer graphene in which an external field saturates the spin relaxation anisotropy.

\end{abstract}

\maketitle

\section{Introduction}
Graphene encapsulated in hBN is emerging as the
long-awaited platform for two-dimensional (2D) spintronics \cite{Han2014:NN, Zutic2004:RMP}.
First generation graphene devices, based on SiO$_2$/Si substrates
\cite{Han2010:PRL, Popinciuc2009:PRB, Han2011:PRL,Han2012:NL, Fan2011:JoP, Gao2014:JoPD, Kang2008:PRB, 
Wang2013:Sc, Maassen2011:PRB, Yang2011:PRL, Tombros2007:Nat},  
show very poor spin transport and ultrafast spin relaxation (SR) 
with spin lifetimes of a few hundred picoseconds. 
In contrast, theory predicts only a few $\mu$eV SOC in 
pristine graphene \cite{Gmitra2009:PRB, Boettger2007:PRB, Konschuh2010:PRB} and 
outstanding spin lifetimes in the nanosecond range 
\cite{Ertler2009:PRB, Pesin2013:NM, Min2006:PRB, Dugaev2011:PRB, Huertas2009:PRL}.
However, due to electron-hole puddles \cite{VanTuan2016:SR, Martin2008:NP}, surface roughness, 
defects and impurities \cite{Sabio2008:PRB, Chen2008:NP} originating from the substrate, 
graphene's SOC can be significantly increased, substantially 
influencing electronic and spin transport properties.
Furthermore, the absence of a marked SR anisotropy 
in these devices \cite{Raes2016:NC, Ringer2018:PRB, Zhu2018:PRB} 
was explained by the presence of magnetic resonant 
scatterers \cite{Irmer2018:PRB, Kochan2014:PRL, Miranda2017:JPCS}.
One attempt of counteracting the substrate's influence is 
to suspend graphene \cite{Guimaraes2012:NL, Du2008:NN, Bao2009:NN}, 
yielding high mobilities but also limited spin transport. 
Therefore the search for new substrates revealed that hBN is the material of interest.

The new generation of graphene devices is based on (hBN)/graphene/hBN stacks
\cite{Roche2015:2DM, Kamalakar2014:APL,Gurram2017:2DM,Guimaraes2014:PRL,
Singh2016:APL,Drogeler2016:NL,Drogeler2017:PSS, Zomer2012:PRB, 
Ingla-Aynes2015:PRB, Drogeler2014:NL, Dean2012:SSC, Dean2010:NN, Wang2017:RSC}, 
which have outstanding transport properties with giant mobilities 
up to 10$^6$ cm$^2$/Vs \cite{Banszerus2015:SA, Petrone2012:NL, Calado2014:APL} 
and record spin lifetimes exceeding 10~ns \cite{Drogeler2016:NL}.
Owing to the improved growth techniques, large scale,
defect free, and smooth interfaces of graphene and hBN \cite{Xue2011:NM, VanTuan2016:SR, 
Mishra2016:C, Tang2013:SR, Arjmandi-Tash2018:JPM, Zomer2011:APL} can be easily produced. 
Especially this second generation of graphene devices
is very important for the realization of spintronics and spin-logic devices 
\cite{Gurram2016:PRB, Mayorov2011:NL, Gurram2017:NC, Gurram2018:PRB, Gurram2017:2DM,
Han2014:NN, Kamalakar2014:APL, Britnell2012:Sc, Wang2017:MTP, Fabian2007:APS, 
Zutic2004:RMP, Lin2013:NL,Lin2014:ACS,Luo2017:NL,Wen2016:PRA,Zutic2006:IBM,Behin2010:NN}.

There is now experimental evidence that in hBN encapsulated bilayer graphene, SR is
due to SOC \cite{Leutenantsmeyer2018:PRL, Xu2018:PRL}. Finally there is a graphene-based
structure in which spins live long (10 ns) and SOC is strong enough, relative to other spin-dependent 
interactions, to play a dominant role and be used for spin manipulation. 
The evidence comes from SR anisotropy. In 2D electron gases in 
semiconductor quantum wells the out-of-plane electron spins
have lifetimes ($\tau_{s,z}$) smaller than in-plane spins ($\tau_{s,x}$), due to the in-plane Rashba
fields \cite{Zutic2004:RMP}. Typically the SR anisotropy ratio $\xi = \tau_{s,z}/\tau_{s,x}$ is 0.5,
reflecting the fact that two spin-orbit field components can flip an out-of-plane spin, 
but only one component can flip the in-plane spin. In contrast, 
as recently predicted \cite{Cummings2017:PRL} and soon experimentally
realized \cite{Leutenantsmeyer2018:PRL, Xu2018:PRL,Zihlmann2018:PRB, Omar2018:PRB}, 
2D materials offer so far unrivaled control over $\xi$. 
It was found that graphene on a transition metal dichalcogenide (TMDC) has $\xi \approx 10$, 
due to the strong valley Zeeman spin-orbit fields, being induced from the TMDC into 
graphene. In this system the spin-orbit fields are relatively large (1 meV \cite{Gmitra2016:PRB}) 
compared to graphene (10 $\mu$eV \cite{Gmitra2009:PRB}), which is also reflected in the 
rather small spin lifetimes of about 10 ps.

On the other hand, the SR anisotropy in encapsulated bilayer graphene 
is also giant ($\xi \approx 10$), but the spin lifetime is three orders of magnitude larger, 
up to 10 ns \cite{Leutenantsmeyer2018:PRL, Xu2018:PRL}. Remarkably, the SR anisotropy $\xi$ 
sharply {\it increases} as a transverse electric field is applied  \cite{Xu2018:PRL} at a fixed doping. 
This is counterintuitive, since the applied field should increase the Rashba field and lower $\tau_{s,z}$. 
The resolution lies in the idiosyncratic spin-orbit band structure of bilayer graphene. 
In the presence of even a moderate electric field, the lowest energy
bands at K split due to SOC, but the splitting does not depend on the field, acquiring the 
intrinsic value of about 24 $\mu$eV \cite{Konschuh2012:PRB}, determined by density-functional theory (DFT).

Since SR anisotropy in mono- and bilayer 
graphene has been a hotly debated issue recently, we ask the following questions.
What are the spin lifetime limits in (hBN)/graphene/hBN heterostructures? 
Does monolayer graphene also have a large SR anisotropy, as shown in hBN encapsulated bilayer graphene? 
Can the anisotropy be tuned electrically? 

Here we focus on monolayer graphene encapsulated in hBN, or placed on a hBN substrate. 
We predict, by DFT calculations and phenomenological modeling, the values of induced spin-orbit
fields, as well as what is the expected SR anisotropy in a variety of potentially realizable structures. 
It is shown (and this should be true for bilayer graphene at low electric fields as well) 
that the anisotropy depends on the actual atomic arrangement of the structures 
and is highly electrically tunable. Unlike in bilayer graphene,
in our systems the anisotropy $\xi$ {\it decreases} with 
increasing electric field, being giant (about 10) at low fields and 
reaching the Rashba limit of 50\% at large fields. The spin lifetimes are expected to be on 
the order of 10 ns, as already seen experimentally,
and also theoretically elaborated for SOC in the tens of $\mu$eV range \cite{VanTuan2016:SR}.

\section{Geometry \& Computational Details}
In order to calculate the electronic band structure of (hBN)/graphene/hBN heterostructures, 
we use a common unit cell for graphene and hBN. 
Therefore, we fix the lattice constant of graphene \cite{Neto2009:RMP} 
to $a = 2.46$~\AA~and change the hBN lattice constant 
from its experimental value \cite{Catellani1987:PRB} 
of $a = 2.504$~\AA~to the graphene one. 
The lattice constants of graphene and hBN differ by less than 2\%, 
justifying our theoretical considerations of commensurate geometries. 
While the small lattice mismatch does lead to moir\'{e} patterns
\cite{Jung2014:PRB,Moon2014:PRB,Argentero2017:NL}, 
the global band structure of local individual stacking configurations
is qualitatively similar \cite{Quhe2012:NPG, Giovannetti2007:PRB}. 
Nevertheless, here we consider all structural arrangements 
for commensurate unit cells, so as to 
get a quantitative feeling for spin-orbit phenomena in a generic experimental setting.

The stacking of graphene on hBN is a crucial point, 
however it was already shown that the configuration
with the lowest energy is, when one C atom is over the B atom and 
the other C atom is over the hollow site of hBN \cite{Giovannetti2007:PRB}.
Before we proceed, we define a terminology to make sense of the 
structural arrangements, used in the following. 
We denote the three relevant sites in hBN as 
the B site (boron), the N site (nitrogen), and the 
H site (hollow position in the center of the hexagon). 
Similarly, we have two graphene sublattices $\alpha$ (C$_{\textrm{A}}$) and 
$\beta$ (C$_{\textrm{B}}$). We call the
energetically most favorable configuration ($\alpha$B, $\beta$H), 
where C$_{\textrm{A}}$ is over Boron, 
and C$_{\textrm{B}}$ is over the hollow site. 
According to this definition we define the other configurations as 
($\alpha$N, $\beta$H) and ($\alpha$N, $\beta$B). 
Due to symmetry, the configurations with interchanged 
C$_{\textrm{A}}$ and C$_{\textrm{B}}$ sublattices give the same results.
The lowest energy interlayer distances between 
graphene and hBN are different for the different stackings \cite{Giovannetti2007:PRB}. 
We include a distance study for all three configurations, in order to  
reveal what are the corresponding lowest energy distances.

\begin{figure}[htb]
 \includegraphics[width=.99\columnwidth]{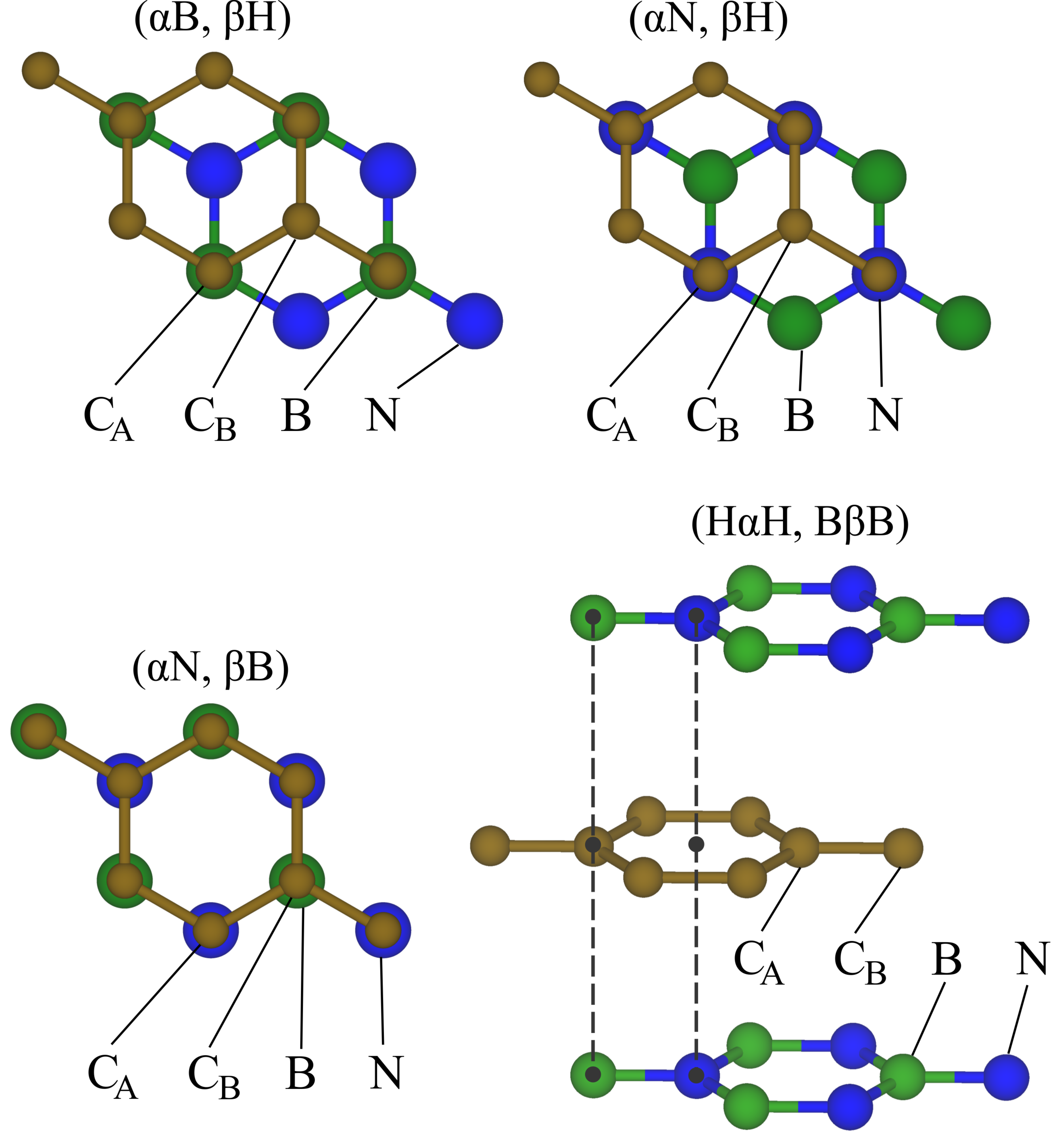}
 \caption{(Color online) Three high-symmetry commensurate stacking configurations 
 of graphene on hBN, ($\alpha$B, $\beta$H), ($\alpha$N, $\beta$H), and ($\alpha$N, $\beta$B) and
 the (H$\alpha$H, B$\beta$B) geometry, as an example of hBN encapsulated graphene. 
 }\label{Fig:stackings}
\end{figure}

In analogy, a stacking sequence of hBN encapsulated 
graphene is then abbreviated as (U$\alpha$V, X$\beta$Y), 
indicating  that the $\alpha$ ($\beta$) sublattice of graphene is sandwiched 
between the U and V (X and Y) sites of top and bottom hBN, 
each of which can take the values \{B, N, H\}. 
It has been shown \cite{Quhe2012:NPG}, that the energetically 
most favorable sandwich structure is (H$\alpha$H, B$\beta$B) in agreement
with our findings here, meaning that $\alpha$ ($\beta$) 
is sandwiched between the two H sites (B sites) of top and bottom hBN.
Interlayer distances, used in the encapsulated geometries, are the ones 
determined by the distance study of the nonencapsulated structures.
In Fig. \ref{Fig:stackings} we show the three commensurate 
stacking configurations of graphene on hBN, 
as well as the (H$\alpha$H, B$\beta$B) geometry, 
as an example of hBN encapsulated graphene.

First-principles calculations are performed with 
full potential linearized augmented plane wave (FLAPW) code 
based on DFT \cite{Hohenberg1964:PRB} and implemented in WIEN2k \cite{Wien2k}. 
Exchange-correlation effects are treated with the 
generalized-gradient approximation (GGA) \cite{Perdew1996:PRL}, including dispersion 
correction \cite{Grimme2010:JCP} and using a $k$-point grid of 
$42\times 42 \times 1$ in the hexagonal Brillouin zone if not specified otherwise. 
The values of the muffin-tin radii we use are $r_{\textrm{C}}=1.34$ for C atom, 
$r_{\textrm{B}}=1.27$ for B atom, and $r_{\textrm{N}}=1.40$ for N atom. 
We use the plane wave cutoff parameter $RK_{\textrm{MAX}} = 9.5$.
In order to avoid interactions between periodic images of our slab geometry, 
we add a vacuum of at least $20$~\AA~in the $z$ direction.

\section{Model Hamiltonian \& Full Band Structure}

The band structure of proximitized graphene can be modeled by 
symmetry-derived Hamiltonians \cite{Kochan2017:PRB}. 
For (hBN)/graphene/hBN heterostructures having $C_{3v}$ symmetry, 
the effective low energy Hamiltonian is
\begin{flalign}
\label{Eq:Hamiltonian}
&\mathcal{H} = \mathcal{H}_{0}+\mathcal{H}_{\Delta}+\mathcal{H}_{\textrm{I}}+\mathcal{H}_{\textrm{R}}+\mathcal{H}_{\textrm{PIA}},\\
&\mathcal{H}_{0} = \hbar v_{\textrm{F}}(\tau k_x \sigma_x - k_y \sigma_y)\otimes s_0, \\
&\mathcal{H}_{\Delta} =\Delta \sigma_z \otimes s_0,\\
&\mathcal{H}_{\textrm{I}} = \tau (\lambda_{\textrm{I}}^\textrm{A} \sigma_{+}+\lambda_{\textrm{I}}^\textrm{B} \sigma_{-})\otimes s_z,\\
&\mathcal{H}_{\textrm{R}} = -\lambda_{\textrm{R}}(\tau \sigma_x \otimes s_y + \sigma_y \otimes s_x),\\
&\mathcal{H}_{\textrm{PIA}} = a(\lambda_{\textrm{PIA}}^\textrm{A} \sigma_{+}-\lambda_{\textrm{PIA}}^\textrm{B} 
\sigma_{-})\otimes (k_x s_y - k_y s_x). 
\end{flalign}
Here $v_{\textrm{F}}$ is the Fermi velocity and the in-plane wave vector 
components $k_x$ and $k_y$ are measured from $\pm$K, 
corresponding to the valley index $\tau = \pm 1$.
The Pauli spin matrices are $s_i$, 
acting on spin space ($\uparrow, \downarrow$), and $\sigma_i$ are pseudospin 
matrices, acting on sublattice space (C$_\textrm{A}$, C$_\textrm{B}$), 
with $i = \{ 0,x,y,z \}$ and $\sigma_{\pm} = \frac{1}{2}(\sigma_z \pm \sigma_0)$. 
The lattice constant is $a = 2.46$~\AA~of pristine graphene and 
the staggered potential gap is $\Delta$.
The parameters $\lambda_{\textrm{I}}^\textrm{A}$ and 
$\lambda_{\textrm{I}}^\textrm{B}$ 
describe the sublattice resolved intrinsic SOC, 
$\lambda_{\textrm{R}}$ stands for the Rashba SOC, 
and $\lambda_{\textrm{PIA}}^\textrm{A}$ and 
$\lambda_{\textrm{PIA}}^\textrm{B}$ are the sublattice 
resolved pseudospin-inversion asymmetry (PIA) SOC parameters. 
The basis states are $|\Psi_{\textrm{A}}, \uparrow\rangle$, 
$|\Psi_{\textrm{A}}, \downarrow\rangle$, $|\Psi_{\textrm{B}}, \uparrow\rangle$, 
and $|\Psi_{\textrm{B}}, \downarrow\rangle$, resulting in 
four eigenvalues $\varepsilon_{1/2}^{\textrm{CB/VB}}$.

\begin{figure}[htb]
 \includegraphics[width=.99\columnwidth]{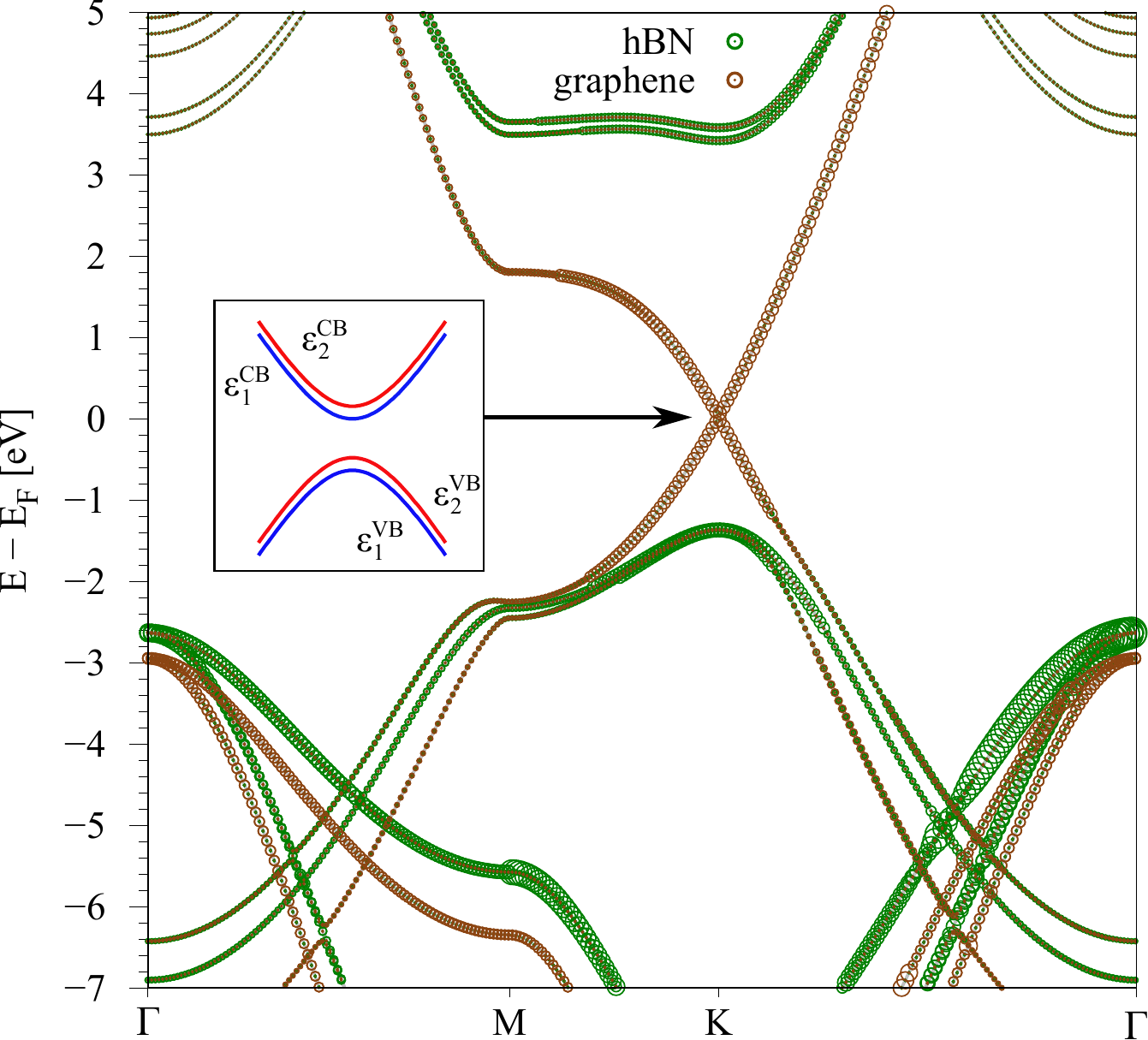}
 \caption{(Color online) Calculated electronic band structure of the 
 (H$\alpha$H, B$\beta$B) geometry, see Fig. \ref{Fig:stackings}.
 The bands of graphene (hBN) are plotted in brown (green). 
 The left inset shows a sketch of the low energy dispersion close to the K point.
  Due to the presence of the substrate, graphene's low energy bands are split 
  into four states $\varepsilon_{1/2}^{\textrm{CB/VB}}$, with a band gap.
 }\label{Fig:bands_HBoCCoHB}
\end{figure}

The calculated band structure of encapsulated graphene 
in the (H$\alpha$H, B$\beta$B) configuration
is shown in Fig. \ref{Fig:bands_HBoCCoHB}, as a representative example 
for all considered geometries. 
Other stacking geometries, as well as graphene on hBN, exhibit similar
band features. The Dirac bands of graphene are located within the hBN band gap. 
In general, the geometries we consider in the following, have broken pseudospin symmetry, 
and a band gap opens in graphene. 
Then, e.g., C$_{\textrm{A}}$ orbitals form the conduction band (CB), 
while C$_{\textrm{B}}$ ones form the valence band (VB).
Further, the low energy bands split into four states $\varepsilon_{1/2}^{\textrm{CB/VB}}$ 
due to SOC and the Rashba effect, see left inset in Fig.~\ref{Fig:bands_HBoCCoHB}.
The general strategy is now to calculate the low energy bands and 
extract the model Hamiltonian parameters best fitting the DFT results, 
for all (hBN)/graphene/hBN geometries.

\section{Graphene on hBN}
In this section we discuss the graphene/hBN heterostructures. 
We show our fit results to the low energy Hamiltonian 
for the different stacking configurations and analyze the influence of the interlayer distance,
between graphene and hBN, on the extracted orbital and SOC model parameters. 
Furthermore, we show and discuss calculated spin-orbit fields. 
Before we turn to the calculation of the SR properties, we
show the tunability of the parameters by applying a transverse electric field for
one specific stacking configuration. 
Finally, we discuss the accuracy of the model, analyze atomic SOC contributions and 
consider an arbitrary but special graphene/hBN stack. 

\subsection{Low energy bands}

\begin{figure}[!htb]
 \includegraphics[width=.98\columnwidth]{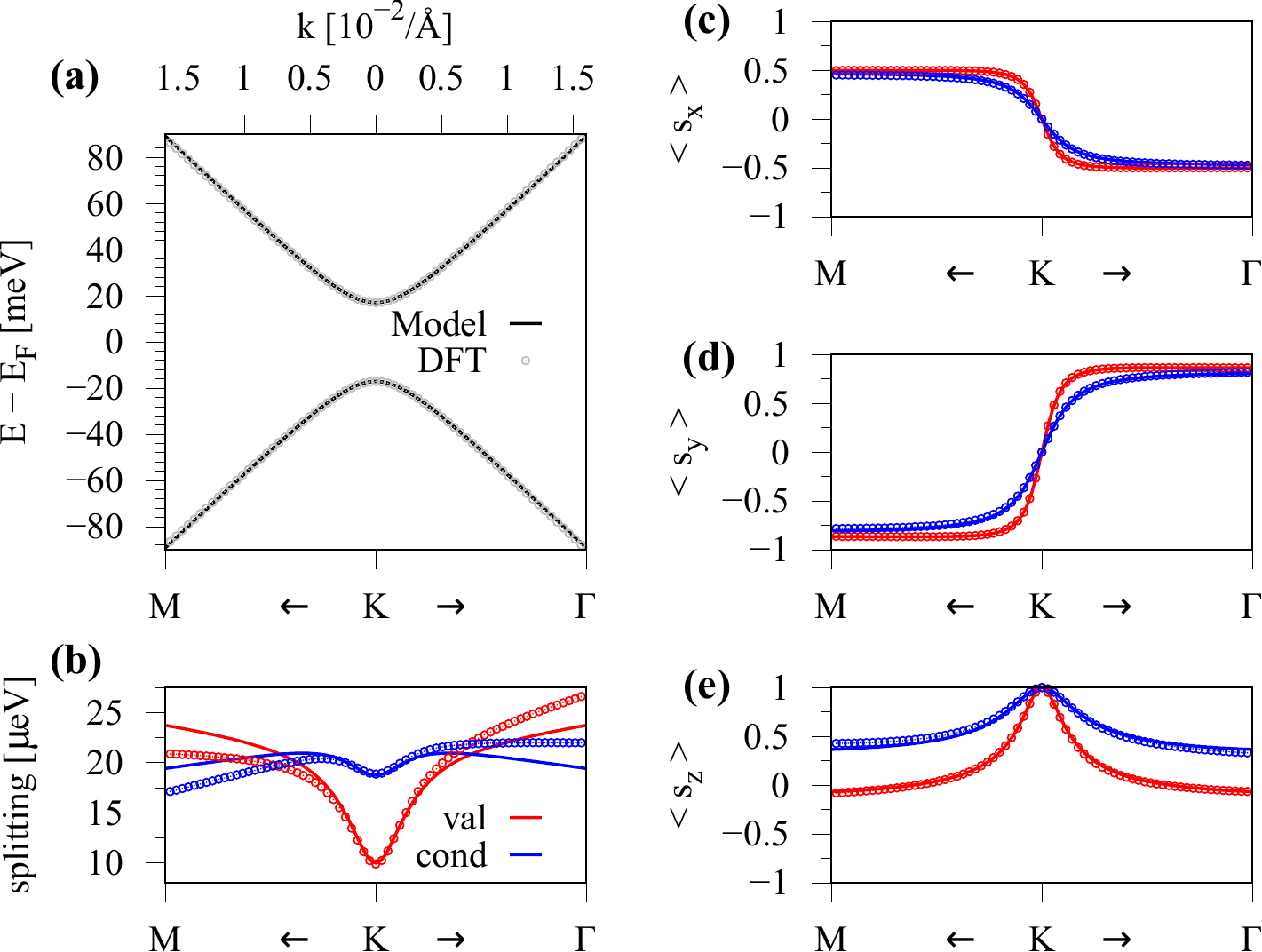}
 \caption{(Color online) Calculated band properties of graphene on hBN in the 
 vicinity of the K point for ($\alpha$B, $\beta$H) configuration
and an interlayer distance of $3.35$~\AA. 
 (a) First-principles band structure (symbols) with a fit to the model Hamiltonian (solid line). 
 (b) The splitting of conduction band $\Delta\textrm{E}_{\textrm{CB}}$ (blue) and valence band 
 $\Delta\textrm{E}_{\textrm{VB}}$ (red) close to the K point and calculated model results. 
 (c)-(e) The spin expectation values of the bands $\varepsilon_{2}^{\textrm{VB}}$ and 
 $\varepsilon_{1}^{\textrm{CB}}$ and comparison to the model results. 
 The fit parameters are given in Tab. \ref{tab:fit_grp_hBN}.  
 }\label{Fig:fitmodel_CCoBH}
\end{figure}
\begin{figure}[!htb]
 \includegraphics[width=.98\columnwidth]{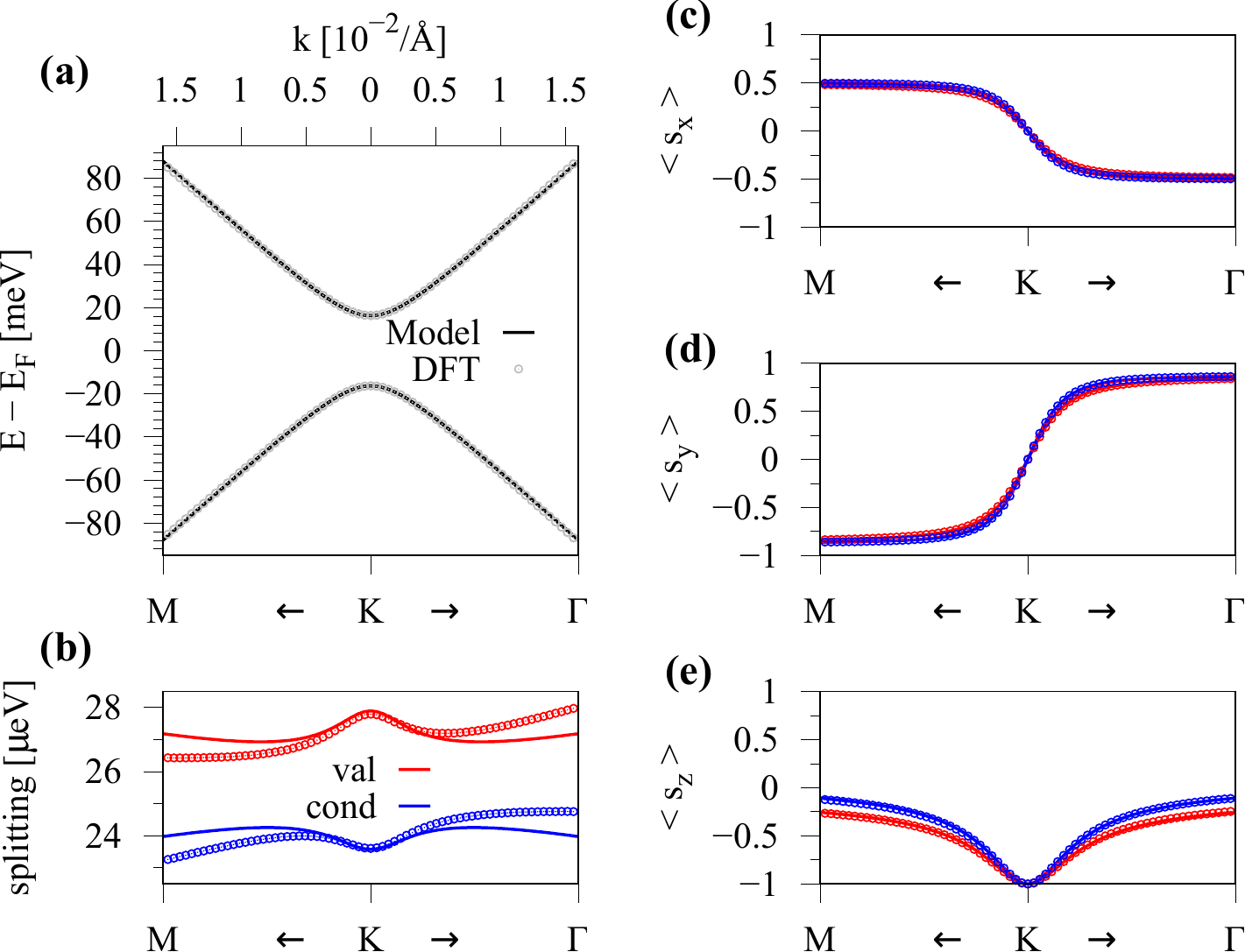}
 \caption{(Color online) Calculated band properties of graphene on hBN in the vicinity 
 of the K point for ($\alpha$N, $\beta$H) configuration 
and an interlayer distance of $3.50$~\AA. 
 (a) First-principles band structure (symbols) with a fit to the model Hamiltonian (solid line). 
 (b) The splitting of conduction band $\Delta\textrm{E}_{\textrm{CB}}$ (blue) and valence band 
 $\Delta\textrm{E}_{\textrm{VB}}$ (red) close to the K point and calculated model results. 
 (c)-(e) The spin expectation values of the bands $\varepsilon_{2}^{\textrm{VB}}$ and 
 $\varepsilon_{1}^{\textrm{CB}}$ and comparison to the model results. 
 The fit parameters are given in Tab. \ref{tab:fit_grp_hBN}.
 }\label{Fig:fitmodel_CCoNH}
\end{figure}
\begin{figure}[!htb]
 \includegraphics[width=.98\columnwidth]{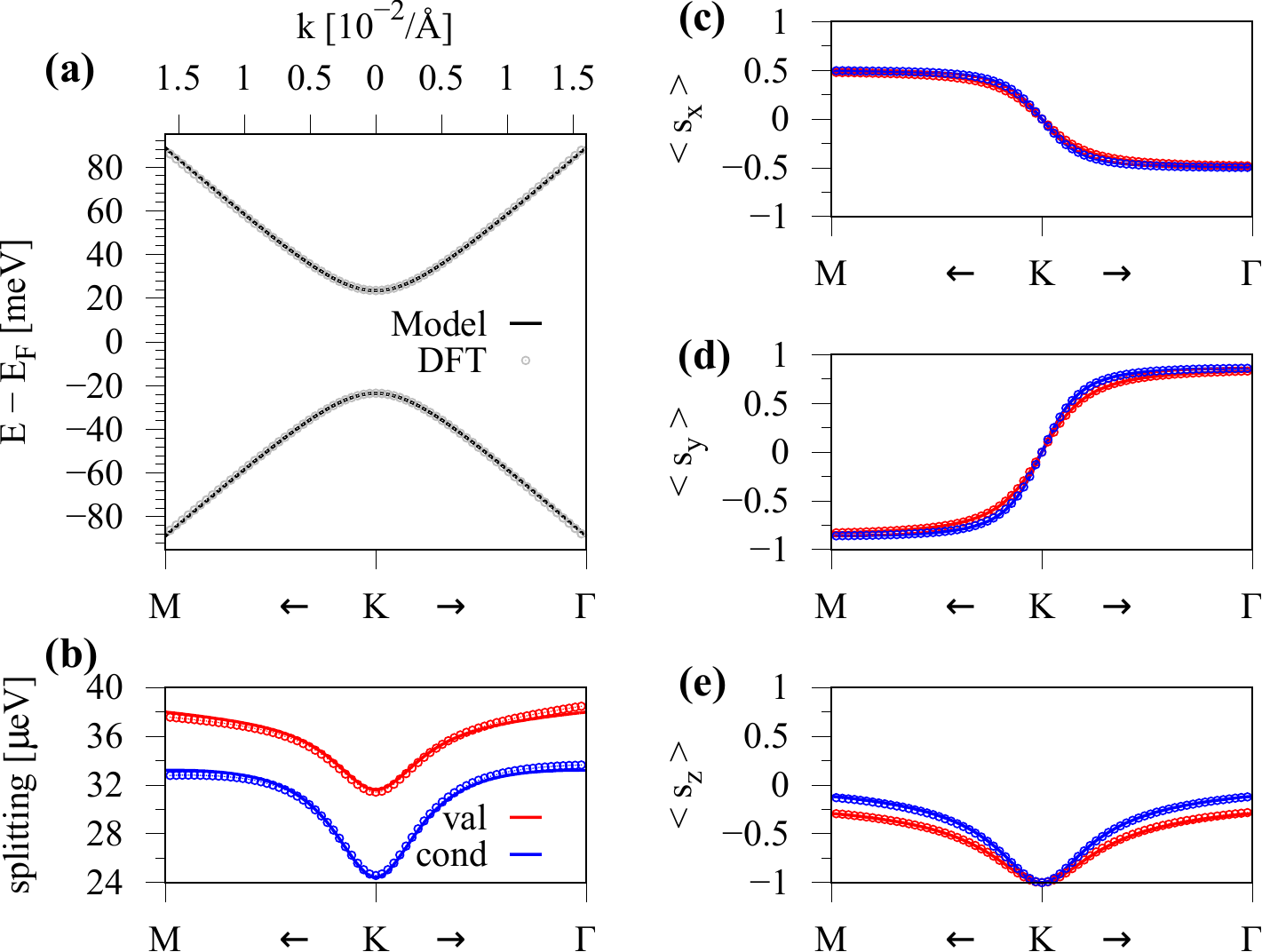}
 \caption{(Color online) Calculated band properties of graphene on hBN in the vicinity 
 of the K point for ($\alpha$N, $\beta$B) configuration 
and an interlayer distance of $3.55$~\AA. 
 (a) First-principles band structure (symbols) with a fit to the model Hamiltonian (solid line). 
 (b) The splitting of conduction band $\Delta\textrm{E}_{\textrm{CB}}$ (blue) and valence band 
 $\Delta\textrm{E}_{\textrm{VB}}$ (red) close to the K point and calculated model results. 
 (c)-(e) The spin expectation values of the bands $\varepsilon_{2}^{\textrm{VB}}$ and 
 $\varepsilon_{1}^{\textrm{CB}}$ and comparison to the model results. 
 The fit parameters are given in Tab. \ref{tab:fit_grp_hBN}.
 }\label{Fig:fitmodel_CCoNB}
\end{figure}

In Fig.~\ref{Fig:fitmodel_CCoBH} we show the calculated low energy 
band structure in the vicinity of the K point with a fit to 
our minimal tight-binding Hamiltonian for the
($\alpha$B, $\beta$H) configuration of graphene on hBN. 
We can see that the orbital band structure is perfectly reproduced by our model, 
see Fig. \ref{Fig:fitmodel_CCoBH}(a), in a quite large energy window around the Fermi level.
The splittings of the bands are shown in Fig. \ref{Fig:fitmodel_CCoBH}(b), 
which are in the $\mu$eV range and are defined as 
$\Delta\textrm{E}_{\textrm{CB}} =\varepsilon_{2}^{\textrm{CB}}-\varepsilon_{1}^{\textrm{CB}}$ and 
$\Delta\textrm{E}_{\textrm{VB}} = \varepsilon_{2}^{\textrm{VB}}-\varepsilon_{1}^{\textrm{VB}}$.
Also the splittings are nicely reproduced by the model, with a maximum discrepancy of about 
10\% compared to the first-principles data. More specifically, 
the splittings are overestimated (underestimated) along
the K-M (K-$\Gamma$) path, by the model. The reason for the discrepancy of the fit
will be explained at a later point.
Finally, Figs. \ref{Fig:fitmodel_CCoBH}(c)-\ref{Fig:fitmodel_CCoBH}(e) show the 
spin expectation values of the bands 
$\varepsilon_{2}^{\textrm{VB}}$ and $\varepsilon_{1}^{\textrm{CB}}$, 
which are in perfect agreement with the model.
The $s_x$ and $s_y$ spin expectation values show a pronounced signature 
of Rashba SOC, with a sign change around the K point. 
The $s_z$ expectation values are maximum at the K point, 
slowly decaying away from it.

In Figs. \ref{Fig:fitmodel_CCoNH} and \ref{Fig:fitmodel_CCoNB} we show the fits 
to the model Hamiltonian for the ($\alpha$N, $\beta$H) and ($\alpha$N, $\beta$B) configurations. 
The overall results look similar to the ($\alpha$B, $\beta$H) configuration.
The orbital band structure, splittings, and spin expectation values are again nicely reproduced by the model.
Compared to the ($\alpha$B, $\beta$H) case, band splittings are even better reproduced 
in these two cases, with a maximum discrepancy of 3\% and 1\%, respectively.
The $s_x$ and $s_y$ spin expectation values again show the characteristic signature of Rashba SOC, 
originating from the broken inversion symmetry of graphene, due to the hBN substrate. 
However, the $s_z$ expectation values, also decaying away from K, 
are opposite compared to the ($\alpha$B, $\beta$H) case. 
We find that our model is very robust and different stacking 
configurations are described by different parameter sets. 
The extracted parameters are given in Tab. \ref{tab:fit_grp_hBN}
for the three commensurate high-symmetry graphene/hBN stacking
configurations, with their corresponding lowest energy distance.

\begin{table*}[!htb]
\begin{ruledtabular}
\begin{tabular}{l  c  c c  c  c   c  c  c}
Configuration & distance [\AA] & $v_{\textrm{F}}/10^5 [\frac{\textrm{m}}{\textrm{s}}]$ & 
$\Delta$~[meV]& $\lambda_{\textrm{R}}~[\mu$eV] & $\lambda_{\textrm{I}}^\textrm{A}~[\mu$eV] &
$\lambda_{\textrm{I}}^\textrm{B}~[\mu$eV] & $\lambda_{\textrm{PIA}}^\textrm{A}~[\mu$eV] & 
$\lambda_{\textrm{PIA}}^\textrm{B}~[\mu$eV] \\
\hline
($\alpha$B, $\beta$H) & 3.35 & 8.308 & -17.08 & 10.65 & 5.00  & 9.37 & 33.58 & 37.57\\
($\alpha$N, $\beta$H) & 3.50 &  8.197 & 16.31 & 12.67 & 11.78 & 13.96 & 4.431 & 26.68\\
($\alpha$N, $\beta$B) & 3.55 & 8.128 & 23.50 & 17.89 & 12.21 & 15.82 & 12.91 & 29.73\\
\hline
average  & 3.47 & 8.211 & 7.577 & 13.74 & 9.66 & 13.05 & 16.97 & 31.33
\end{tabular}
\end{ruledtabular}
\caption{\label{tab:fit_grp_hBN} Fit parameters for the three graphene/hBN stacks at their 
energetically most favorable distances. The Fermi velocity $v_{\textrm{F}}$, gap parameter $\Delta$, 
 Rashba SOC parameter $\lambda_{\textrm{R}}$, 
 intrinsic SOC parameters $\lambda_{\textrm{I}}^\textrm{A}$ and $\lambda_{\textrm{I}}^\textrm{B}$, 
 and PIA SOC parameters $\lambda_{\textrm{PIA}}^\textrm{A}$ and $\lambda_{\textrm{PIA}}^\textrm{B}$.  
 In the last row we average over the configurations for each parameter. }
\end{table*}

\subsection{Distance study}

\begin{figure}[htb]
 \includegraphics[width=.99\columnwidth]{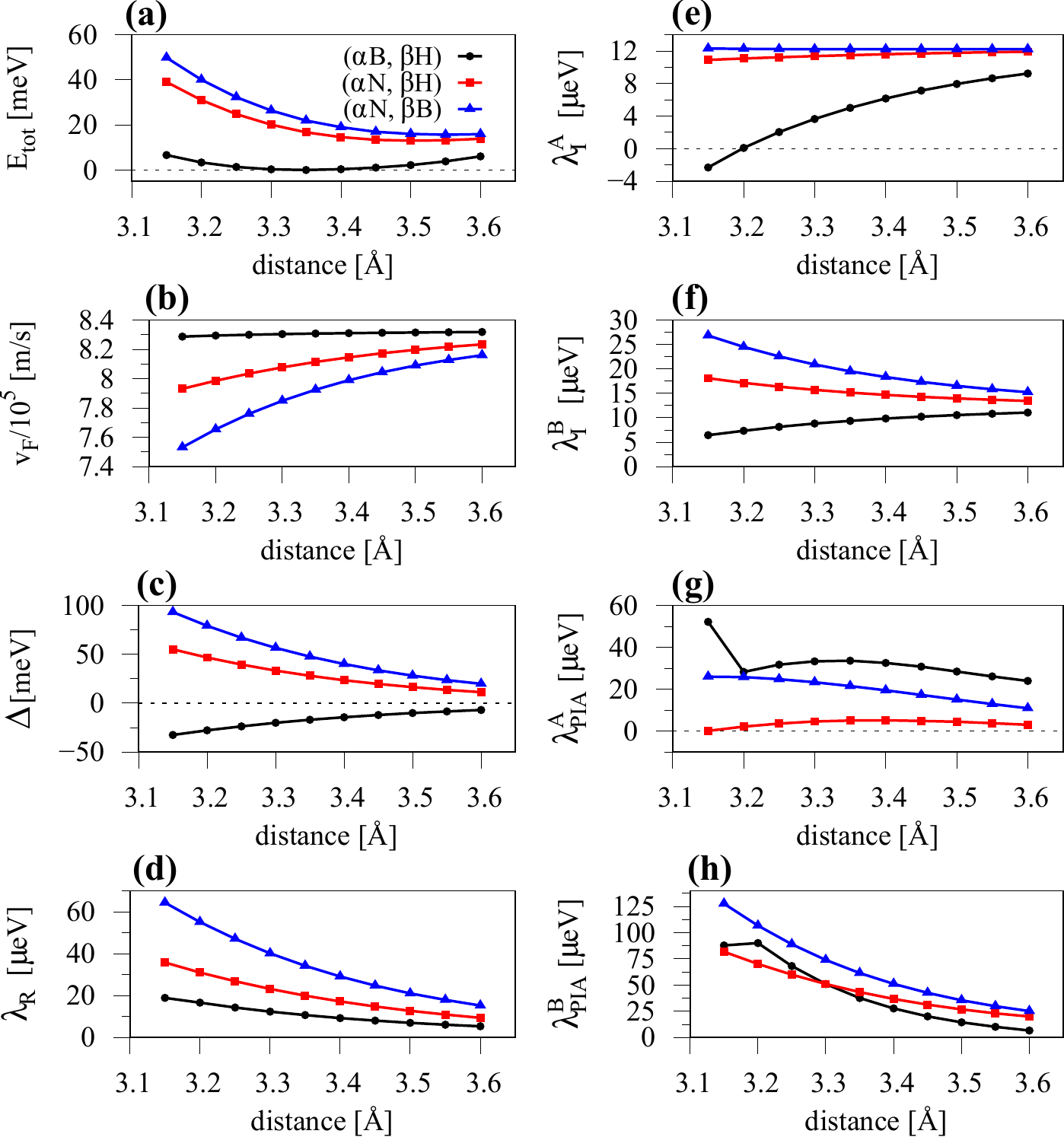}
 \caption{(Color online) Fit parameters as a function of interlayer distance between graphene and hBN for the 
 three different stacking configurations. (a) Total energy, (b) the Fermi velocity $v_{\textrm{F}}$ (c) gap parameter $\Delta$, 
 (d) Rashba SOC parameter $\lambda_{\textrm{R}}$, (e) intrinsic SOC parameter $\lambda_{\textrm{I}}^\textrm{A}$ for sublattice A, 
 (f) intrinsic SOC parameter $\lambda_{\textrm{I}}^\textrm{B}$ for sublattice B, (g) PIA SOC parameter 
 $\lambda_{\textrm{PIA}}^\textrm{A}$ for sublattice A, and (h) PIA SOC parameter $\lambda_{\textrm{PIA}}^\textrm{B}$ 
 for sublattice B.
 }\label{Fig:distance_grp_hBN}
\end{figure}
One has to mention that different stacking configurations lead to different interlayer distances 
between graphene and hBN, when minimizing the total energy of the individual geometries. 
In Fig. \ref{Fig:distance_grp_hBN} we show the fit parameters, as a 
function of the distance between graphene and hBN, 
for the three stacking configurations. 
We find that the total energy is lowest for the ($\alpha$B, $\beta$H) 
configuration with an interlayer distance of $3.35$~\AA, see Fig. \ref{Fig:distance_grp_hBN}(a).
The lowest energies for the ($\alpha$N, $\beta$H) and ($\alpha$N, $\beta$B) 
configurations are obtained at distances of $3.50$~\AA~and $3.55$~\AA. 
The Fermi velocity $v_{\textrm{F}}$, see Fig. \ref{Fig:distance_grp_hBN}(b), 
which reflects the nearest neighbor hopping strength via
$t = \frac{2\hbar v_{\textrm{F}}}{\sqrt{3}a}$, grows as a function of distance, 
especially for the ($\alpha$N, $\beta$H) and ($\alpha$N, $\beta$B) configurations. 
In contrast to that, the gap parameter $\Delta$ decreases with distance, 
in agreement with literature \cite{Giovannetti2007:PRB}. 
When moving the graphene away from the substrate,
the sublattice symmetry breaking reduces and the gap decreases.

One very important observation is that the gap parameter $\Delta$ of the ($\alpha$B, $\beta$H) configuration
is opposite in sign compared to the other configurations, 
as seen in a moir\'{e} pattern \cite{Miller2010:NP, Hunt2013:SC, Kindermann2012:PRB}. 
In the ($\alpha$B, $\beta$H) configuration, the C$_{\textrm{A}}$ sublattice is over the boron.
Sublattice C$_{\textrm{A}}$ forms, in this case, the VB which is why 
we need a negative value of $\Delta$ in the model, 
to match the sublattice character of the DFT results. 
In contrast, the other configurations have the C$_{\textrm{A}}$ 
sublattice over the nitrogen, which then forms the CB, 
leading to a positive value of $\Delta$.
This also explains why the $s_z$ spin expectation values for different configurations are different, 
compare Figs. \ref{Fig:fitmodel_CCoBH}(e) and \ref{Fig:fitmodel_CCoNH}(e).
In a moir\'{e} pattern geometry, with micrometer size flakes of graphene and hBN, 
all of these local stacking configurations appear simultaneously. 
Consequently, there can be a local stacking geometry where the orbital gap closes, 
appearing when the two sublattices feel the same surrounding potential. 
We will calculate and discuss such a situation at a later point, for a 
certain choice of stacking.

The Rashba SOC parameter, see Fig. \ref{Fig:distance_grp_hBN}(d), also decreases with distance. 
When the distance between graphene and hBN approaches infinity, 
the inversion symmetry of graphene is restored and the Rashba SOC parameter vanishes. 
The two intrinsic SOC parameters $\lambda_{\textrm{I}}^\textrm{A}$ and  $\lambda_{\textrm{I}}^\textrm{B}$ approach the intrinsic 
SOC of $12~\mu$eV of pristine graphene \cite{Gmitra2009:PRB}, 
as we increase the distance, see Figs. \ref{Fig:distance_grp_hBN}(e) and \ref{Fig:distance_grp_hBN}(f).  
Finally, we find that the two PIA SOC parameters $\lambda_{\textrm{PIA}}^\textrm{A}$ and 
$\lambda_{\textrm{PIA}}^\textrm{B}$, see Figs. \ref{Fig:distance_grp_hBN}(g) and \ref{Fig:distance_grp_hBN}(h), 
also decrease with distance.
Overall, as expected, we restore the pristine graphene properties, 
as the interlayer distance gradually increases.

\subsection{Spin-Orbit fields}
\begin{figure}[htb]
 \includegraphics[width=.99\columnwidth]{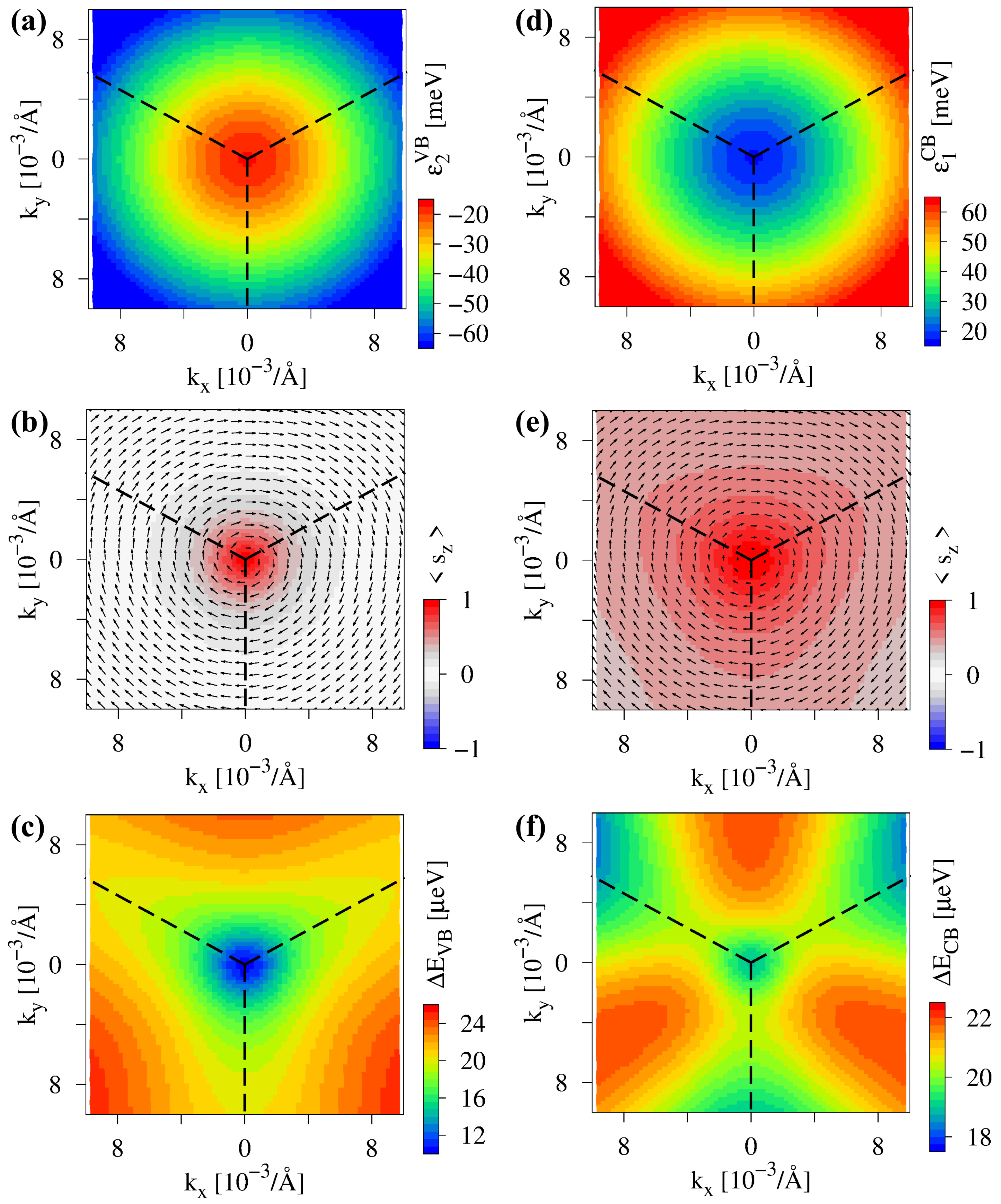}
 \caption{(Color online) Calculated low energy dispersion of 
 graphene on hBN around the K point for ($\alpha$B, $\beta$H) configuration
and an interlayer distance of $3.35$~\AA. 
 (a) 2D map of the energy of the valence band $\varepsilon_{2}^{\textrm{VB}}$, with the corresponding
 spin texture of the band shown in (b) and the splitting of the valence band  
 $\Delta\textrm{E}_{\textrm{VB}} = \varepsilon_{2}^{\textrm{VB}}-\varepsilon_{1}^{\textrm{VB}}$ shown in (c). 
 (d)-(f) The same as (a)-(c), but for conduction band $\varepsilon_{1}^{\textrm{CB}}$ and conduction band splitting
 $\Delta\textrm{E}_{\textrm{CB}} = \varepsilon_{2}^{\textrm{CB}}-\varepsilon_{1}^{\textrm{CB}}$. The dashed lines
 show the edges of the Brillouin zone with the K point at the center. 
 }\label{Fig:spinmap_CCoBH}
\end{figure}
\begin{figure}[htb]
 \includegraphics[width=.99\columnwidth]{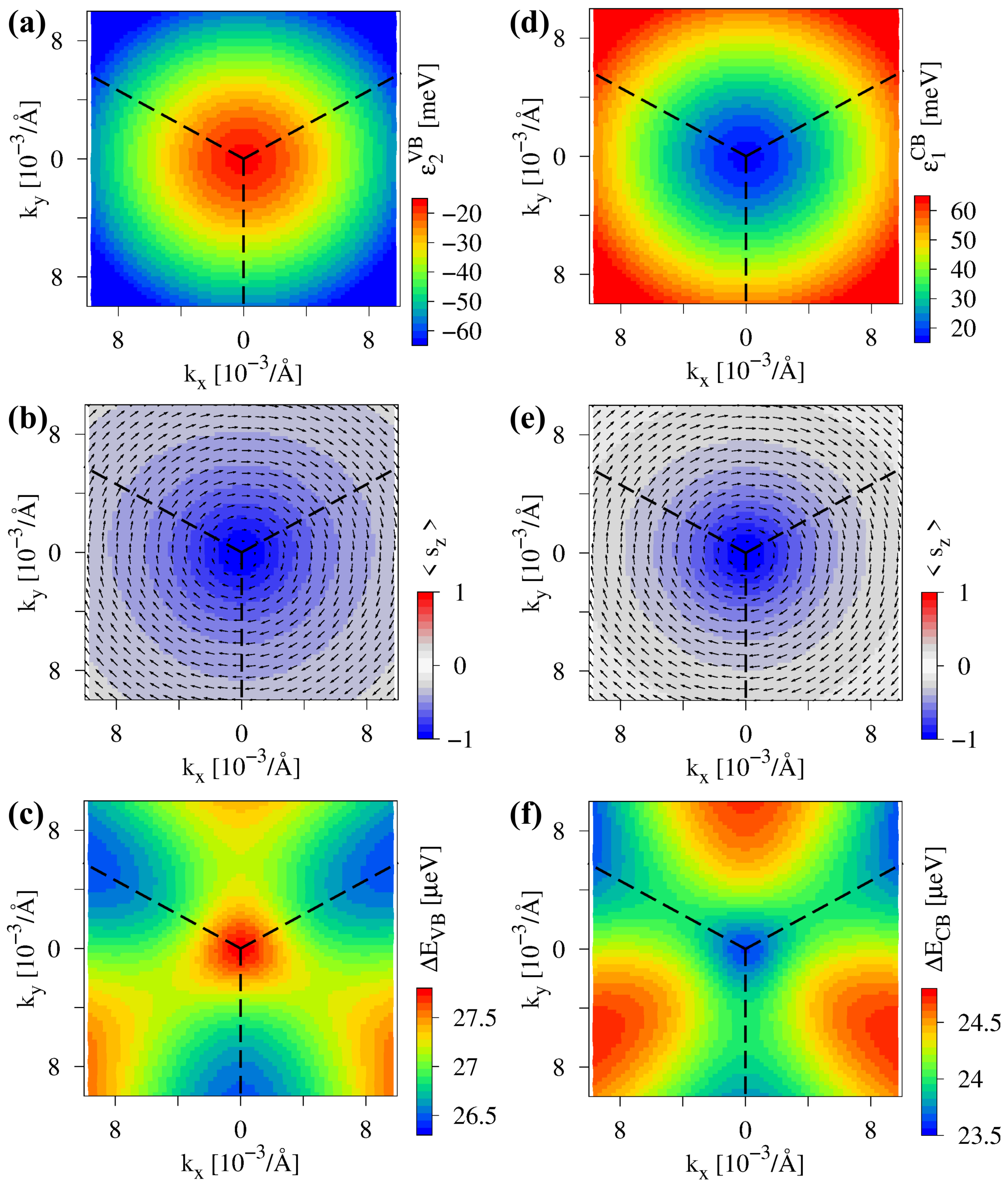}
 \caption{(Color online) Calculated low energy dispersion of graphene on hBN around the 
 K point for ($\alpha$N, $\beta$H) configuration
and an interlayer distance of $3.50$~\AA. 
 (a) 2D map of the energy of the valence band $\varepsilon_{2}^{\textrm{VB}}$, with the corresponding
 spin texture of the band shown in (b) and the splitting of the valence band  
 $\Delta\textrm{E}_{\textrm{VB}} = \varepsilon_{2}^{\textrm{VB}}-\varepsilon_{1}^{\textrm{VB}}$ shown in (c). 
 (d)-(f) The same as (a)-(c), but for conduction band $\varepsilon_{1}^{\textrm{CB}}$ and conduction band splitting
 $\Delta\textrm{E}_{\textrm{CB}} = \varepsilon_{2}^{\textrm{CB}}-\varepsilon_{1}^{\textrm{CB}}$. The dashed lines
 show the edges of the Brillouin zone with the K point at the center. 
 }\label{Fig:spinmap_CCoNH}
\end{figure}
\begin{figure}[htb]
 \includegraphics[width=.99\columnwidth]{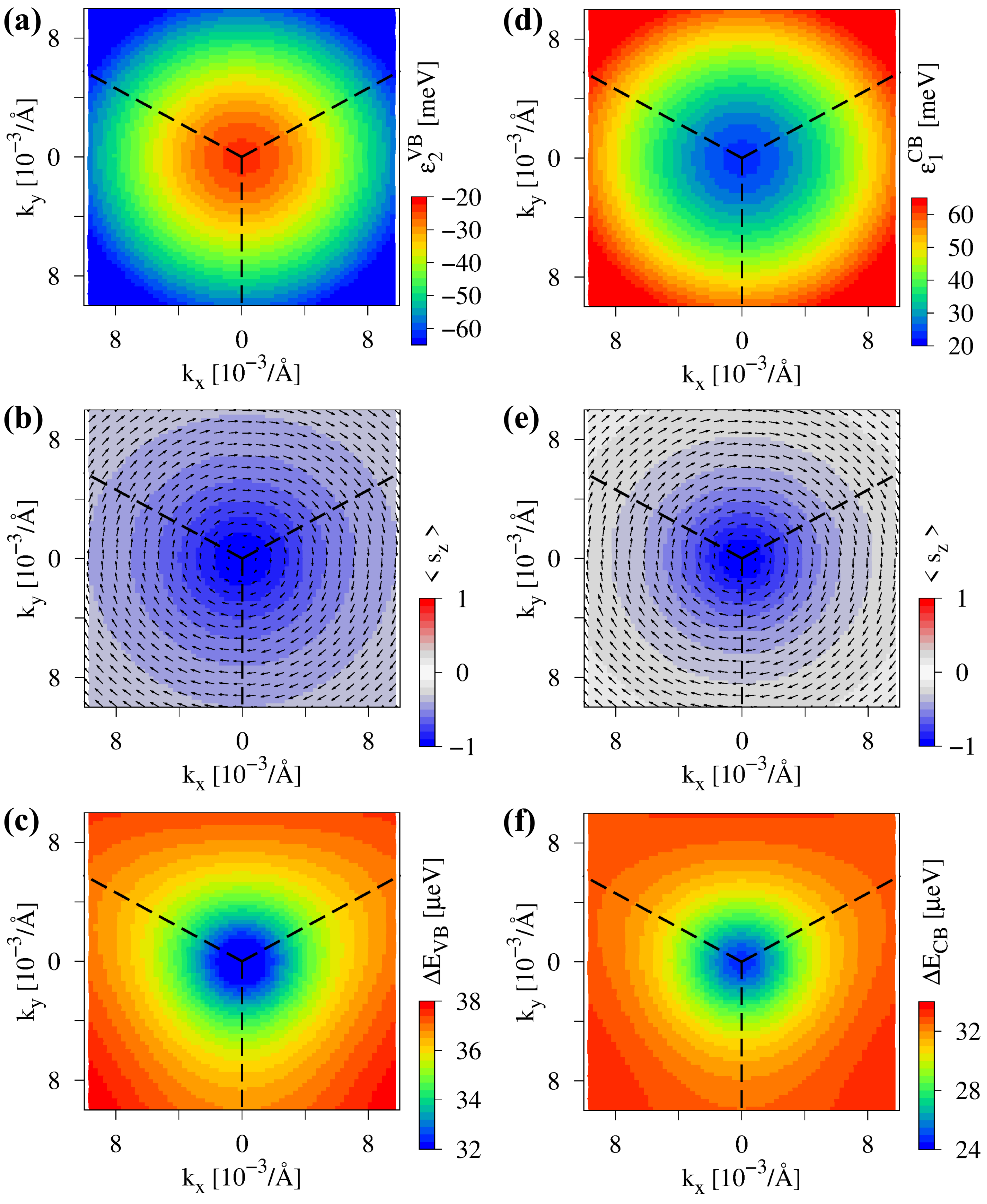}
 \caption{(Color online) Calculated low energy dispersion of graphene 
 on hBN around the K point for ($\alpha$N, $\beta$B) configuration
and an interlayer distance of $3.55$~\AA. 
 (a) 2D map of the energy of the valence band $\varepsilon_{2}^{\textrm{VB}}$, with the corresponding
 spin texture of the band shown in (b) and the splitting of the valence band  
 $\Delta\textrm{E}_{\textrm{VB}} = \varepsilon_{2}^{\textrm{VB}}-\varepsilon_{1}^{\textrm{VB}}$ shown in (c). 
 (d)-(f) The same as (a)-(c), but for conduction band $\varepsilon_{1}^{\textrm{CB}}$ and conduction band splitting
 $\Delta\textrm{E}_{\textrm{CB}} = \varepsilon_{2}^{\textrm{CB}}-\varepsilon_{1}^{\textrm{CB}}$. The dashed lines
 show the edges of the Brillouin zone with the K point at the center. 
 }\label{Fig:spinmap_CCoNB}
\end{figure}

In Fig. \ref{Fig:spinmap_CCoBH} we show the calculated dispersion as a 2D map 
in $k_x$-$k_y$ plane in the vicinity of the K point for the ($\alpha$B, $\beta$H) configuration.
The energies of the bands, $\varepsilon_{2}^{\textrm{VB}}$ and $\varepsilon_{1}^{\textrm{CB}}$,
do not show any trigonal warping, see Figs. \ref{Fig:spinmap_CCoBH}(a) and \ref{Fig:spinmap_CCoBH}(d). 
However, already the spin texture shows that trigonal warping
is present with a very pronounced Rashba spin-orbit field, rotating in a clockwise direction, 
see Figs. \ref{Fig:spinmap_CCoBH}(b) and \ref{Fig:spinmap_CCoBH}(e), 
as expected from the inversion symmetry breaking by the hBN substrate. 
In contrast to the CB, the VB $s_z$ spin expectation 
value strongly decays away from the K point. 
A pronounced threefold symmetry is observed in the spin splittings 
$\Delta\textrm{E}_{\textrm{VB}}$ and $\Delta\textrm{E}_{\textrm{CB}}$, 
see Figs. \ref{Fig:spinmap_CCoBH}(c) and \ref{Fig:spinmap_CCoBH}(f).
Along the K-$\Gamma$ path, the Dirac bands are more split than along the K-M path. 
As we have seen, our model Hamiltonian agrees very well on a qualitative 
level for this case, see Fig. \ref{Fig:fitmodel_CCoBH}, 
however the band splittings cannot be fully recovered.
The reason will be explained in the last subsection. 
In Figs. \ref{Fig:spinmap_CCoNH} and \ref{Fig:spinmap_CCoNB} we show the calculated dispersion 
as a 2D map in the $k_x$-$k_y$ plane in the vicinity of the K point
for the ($\alpha$N, $\beta$H) and ($\alpha$N, $\beta$B) configurations. 
The overall trigonal symmetry features remain and are very similar to the ($\alpha$B, $\beta$H) configuration.
Especially for the ($\alpha$N, $\beta$B) configuration, 
only weak trigonal symmetry, around the K point, can be observed.

\subsection{Transverse electric field}

In experiment gating is required to tune the Fermi level towards the charge neutrality point. 
By using top and back gate electrodes, one can tune the doping level and simultaneously apply an 
electric field across a heterostructure.
Thereby the transverse electric field can influence electronic and
spin-orbit properties of graphene, especially the Rashba SOC \cite{Gmitra2009:PRB}.
We consider the lowest energy configuration ($\alpha$B, $\beta$H) 
for graphene on hBN and apply a transverse electric field, 
which is modeled by a zigzag potential, across the heterostructure. 

\begin{figure}[htb]
 \includegraphics[width=.99\columnwidth]{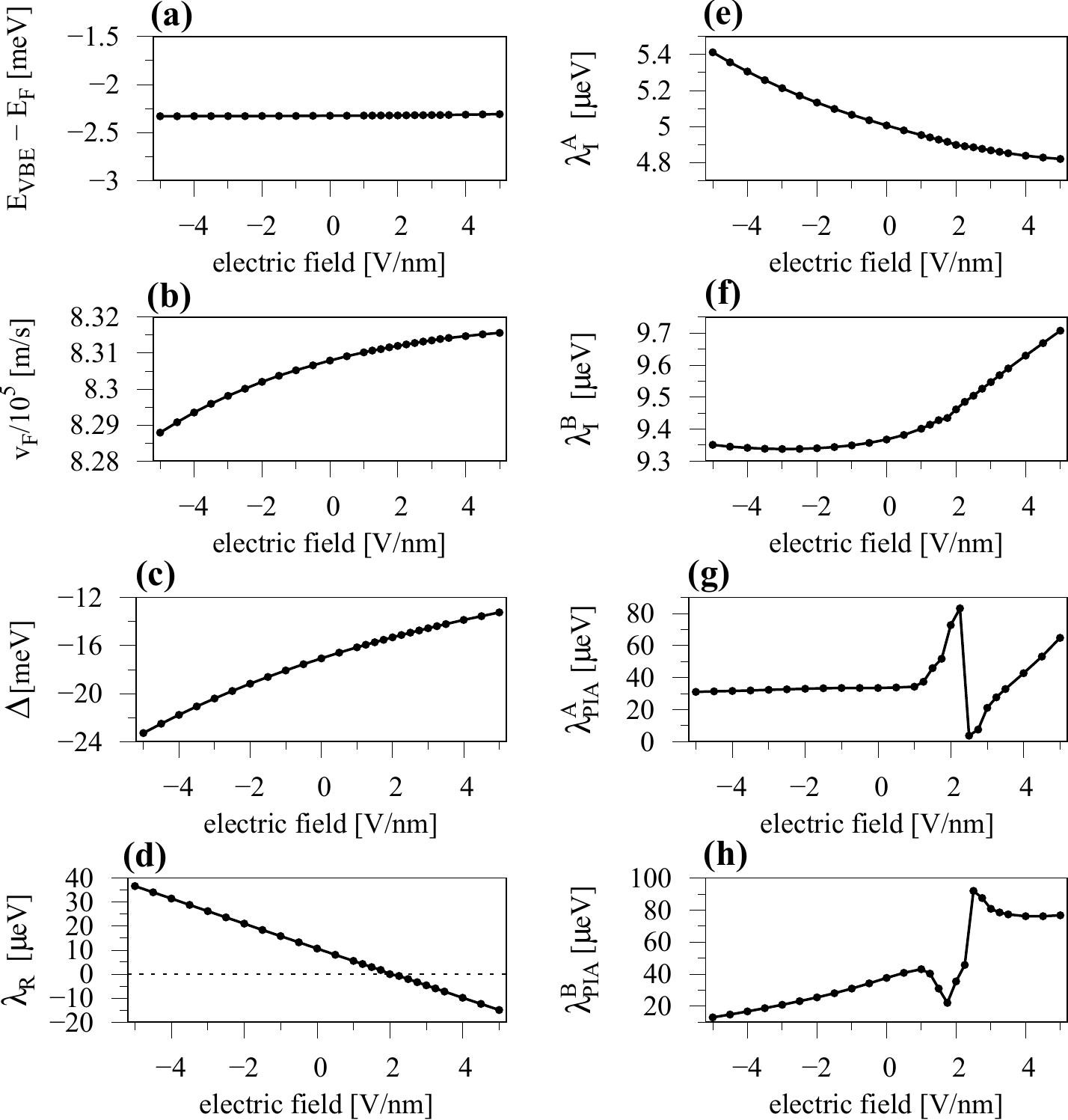}
 \caption{(Color online)  Fit parameters as a function of the applied transverse electric 
 field for the ($\alpha$B, $\beta$H) configuration. 
 (a) Valence band edge with respect to the Fermi level, (b) the Fermi velocity $v_{\textrm{F}}$, (c) gap parameter $\Delta$, 
 (d) Rashba SOC parameter $\lambda_{\textrm{R}}$, (e) intrinsic SOC parameter $\lambda_{\textrm{I}}^\textrm{A}$ for sublattice A, 
 (f) intrinsic SOC parameter $\lambda_{\textrm{I}}^\textrm{B}$ for sublattice B, (g) PIA SOC parameter 
 $\lambda_{\textrm{PIA}}^\textrm{A}$ for sublattice A, and (h) PIA SOC parameter $\lambda_{\textrm{PIA}}^\textrm{B}$ 
 for sublattice B.
 }\label{Fig:Efield_CCoBH}
\end{figure}

For every magnitude of the field we calculate the low energy 
band structure and fit it to the model Hamiltonian. 
In Fig. \ref{Fig:Efield_CCoBH} we show the fit parameters for ($\alpha$B, $\beta$H) configuration
as a function of external electric field. Indeed, we can tune most of 
the parameters. 
The Fermi velocity $v_{\textrm{F}}$, as well as intrinsic SOC parameters 
$\lambda_{\textrm{I}}^\textrm{A}$ and $\lambda_{\textrm{I}}^\textrm{B}$, are barely affected.
However, the field can tune the orbital gap, Rashba and PIA SOC parameters.
Especially the Rashba parameter can be tuned over a wide range, even from positive to negative values, 
with the transition at around 2 V/nm. 
Tuning the Rashba SOC parameter, from a positive to a negative value, also allows us to change
the rotation direction of the spin-orbit fields, see Figs. \ref{Fig:spinmap_CCoBH}(b) and \ref{Fig:spinmap_CCoBH}(e).
Most importantly we can tune the Rashba SOC from a finite value to zero. 
Consequently, we can control the strength of the in-plane spin-orbit field, dictated by Rashba SOC,
which will significantly influence spin transport and SR properties. 
Another feature we notice is that around 2 V/nm, the PIA SOC parameters 
are not changing very smoothly with applied field,
which is connected with the transition of the Rashba SOC through zero.

\subsection{Spin relaxation anisotropy}
Since the low energy Hamiltonian $\mathcal{H}$ can nicely reproduce the
dispersion around the K point, we can use it together with our fit parameters to calculate SR times.
We calculate, for a very dense $k$ grid in the vicinity of the K point, 
the energy spectrum and spin expectation values for the Dirac bands from our model.  
To calculate the SR time, we define the spin-orbit field components $\omega_{k,\textrm{i}}$ as \cite{Fabian2007:APS} 
\begin{equation}
\omega_{k,\textrm{i}} = \frac{\Delta\textrm{E}_{k}}{\hbar}  \cdot \frac{s_{k,\textrm{i}}}{s_k} ,
\end{equation}
where $k$ is the momentum and $s_{k,\textrm{i}}$ are the spin expectation values 
along the direction $\textrm{i} = \{\textrm{x, y, z}\}$. 
The energy splitting of the Dirac bands is $\Delta\textrm{E}_{k}$ 
and $s_k~=~\sqrt{s_{k,\textrm{x}}^2+s_{k,\textrm{y}}^2+s_{k,\textrm{z}}^2}$ 
is the absolute value of the spin. 
By that we obtain at each $k$ point the spin-orbit vector field.
Following the derivation of Refs. \onlinecite{Cummings2017:PRL, Garcia2018:CSR}, we then 
calculate the SR times as follows
\begin{eqnarray}
&\tau_{s,\textrm{x}}^{-1}(\textrm{E}) =  
\tau_p\cdot\langle\omega_{k,\textrm{y}}^2\rangle +\tau_{iv}\cdot 
\langle\omega_{k,\textrm{z}}^2\rangle,\\
&\tau_{s,\textrm{y}}^{-1}(\textrm{E}) =  
\tau_p\cdot\langle\omega_{k,\textrm{x}}^2\rangle +\tau_{iv}\cdot 
\langle\omega_{k,\textrm{z}}^2\rangle,\\
&\tau_{s,\textrm{z}}^{-1}(\textrm{E}) =  
\tau_p\cdot\langle\omega_{k,\textrm{x}}^2+\omega_{k,\textrm{y}}^2\rangle.
\end{eqnarray}
The average $\langle \cdot \rangle$ is taken over all $k$ points that have the same constant energy $E$.
The momentum relaxation time is $\tau_p$ and $\tau_{iv}$ is the intervalley scattering time.
For the calculation of the averages $\langle \cdot \rangle$ we use energy steps of 
$100~\mu$eV with a smearing of $\pm50~\mu$eV, corresponding to a temperature of $0.58$~K. 
Measurements \cite{Drogeler2016:NL, Gurram2017:2DM, Guimaraes2014:PRL, Singh2016:APL} 
provide SR lengths of 
$\lambda_s \approx 20~\mu$m, SR times of $\tau_s \approx 8$~ns, 
and spin diffusion constants of $D_s \approx 0.04~\frac{\textrm{m}^2}{\textrm{s}}$.
With the relation $\lambda_s = \sqrt{\tau_s D_s}$ 
and using that the spin diffusion constant is roughly equal to the charge diffusion constant
$D_s \approx D_c = \frac{1}{2}v_{\textrm{F}}^2\tau_p$ and 
$v_{\textrm{F}}\approx 8\times 10^{5}~\frac{\textrm{m}}{\textrm{s}}$,
we get $\tau_p = 125$~fs, which we use in the calculations. 
The value for $\tau_p$ is reasonable, assuming ultraclean samples.

\begin{figure}[htb]
 \includegraphics[width=.98\columnwidth]{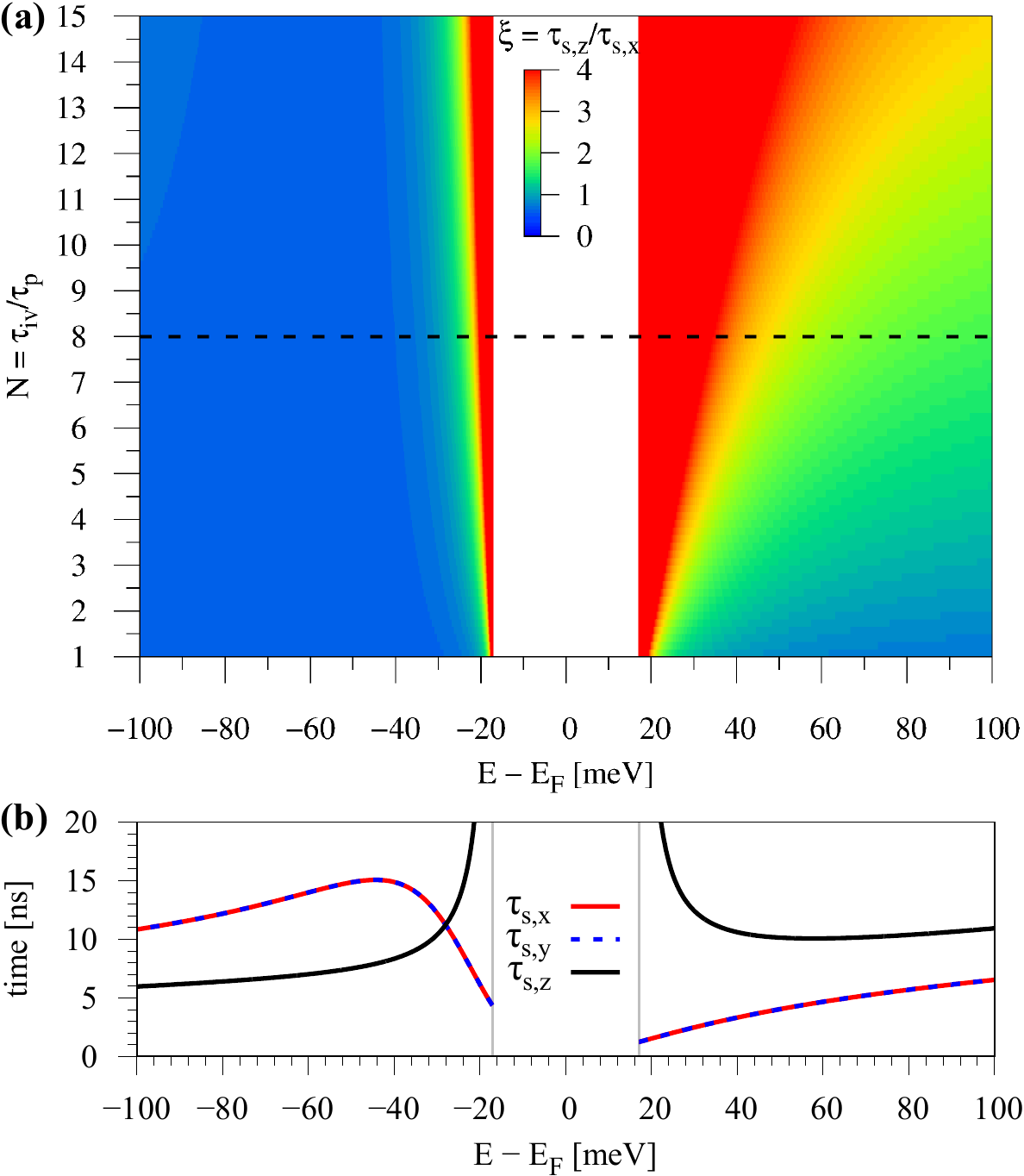}
 \caption{(Color online) Calculated SR times and anisotropies 
 for ($\alpha$B, $\beta$H) configuration. (a) Colormap of 
 the SR anisotropy $\xi = \tau_{s,\textrm{z}}/\tau_{s,\textrm{x}}$ 
 as a function of $N = \tau_{iv}/\tau_p$ and the energy. 
 (b) Individual SR times as a function of energy corresponding to the 
 dashed line in (a) with $\tau_p = 125$~fs and $\tau_{iv}= 8\cdot\tau_p$. 
The gray lines indicate the band edges. 
 }\label{Fig:SRT_CCoBH}
\end{figure}

Since intervalley scattering times are hard to estimate from experiments, 
we consider it variable, 
$\tau_{iv} = N\cdot\tau_p$ with $N = \{1,...,15\}$, for our calculations. 
By that we obtain the SR time as a function of the energy, 
for spins along the $x$, $y$, and $z$ direction for each ratio $N = \tau_{iv}/\tau_p$.
More interesting than the individual SR times
is the SR anisotropy $\xi = \tau_{s,\textrm{z}}/\tau_{s,\textrm{x}}$, 
a measurable fingerprint of the SOC of the system.

\begin{figure}[htb]
 \includegraphics[width=.95\columnwidth]{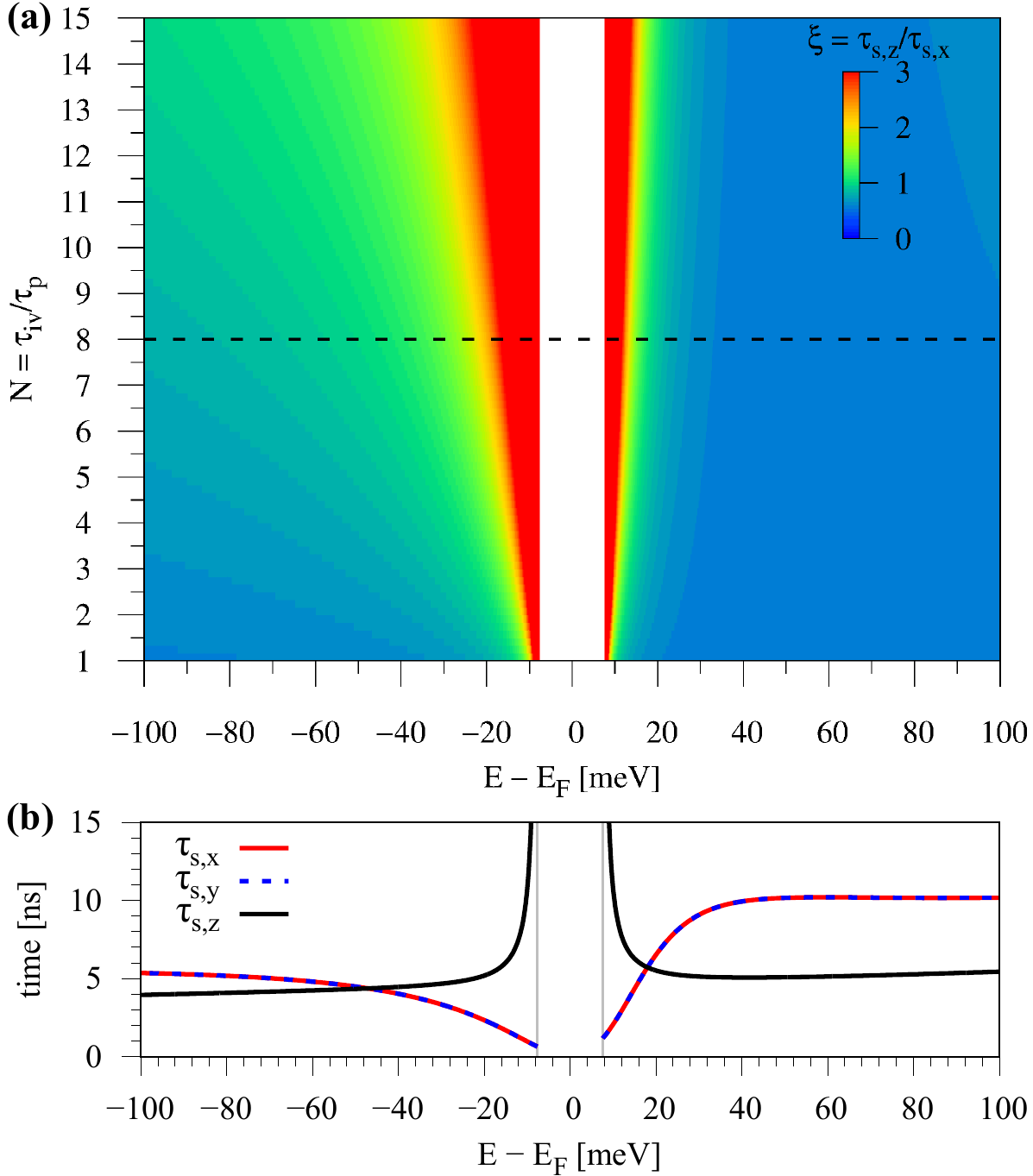}
 \caption{(Color online) Calculated SR times and anisotropies for graphene on hBN. Here we use the 
 averaged parameters of the graphene/hBN heterostructures given in the main text. (a) Colormap of 
 the SR anisotropy $\xi = \tau_{s,\textrm{z}}/\tau_{s,\textrm{x}}$ 
 as a function of $N = \tau_{iv}/\tau_p$ and the energy. (b) Individual SR times as a function of energy 
corresponding to the dashed line in (a) with $\tau_p = 125$~fs and $\tau_{iv}= 8\cdot\tau_p$. 
The gray lines indicate the band edges. 
 }\label{Fig:SRT_grp_hBN}
\end{figure}

We show a colormap of the calculated anisotropy $\xi$ as a function of $N$ and the energy 
for the ($\alpha$B, $\beta$H) configuration in Fig. \ref{Fig:SRT_CCoBH}(a). 
Within the band gap of $\pm 17$~meV, of course no states are available 
and SR times cannot be calculated, because the smearing we use is only $0.58$~K. 
For holes we find that the anisotropy is $\xi \approx \frac{1}{2}$, the Rashba limit, 
as soon as we are below $-20$~meV from the valence band edge, for each ratio $N$. 
For electrons the situation is completely different and the anisotropy can get very large, 
even $20$~meV away from the conduction band edge.
We also find that, independent of $N$, the anisotropy is largest close 
to the band edges, which would correspond to the charge neutrality point in experiment. 
In Fig. \ref{Fig:SRT_CCoBH}(b) we show the individual SR times as a function of energy, 
corresponding to $N = 8$. We find SR times of around 10 ns, 
consistent with measurements \cite{Drogeler2016:NL}. 
However, we have to keep in mind that Fig. \ref{Fig:SRT_CCoBH} 
is only valid for a certain stacking configuration, the ($\alpha$B, $\beta$H) one, of graphene on hBN.

In experiment one expects, that electrons traveling through graphene on a hBN substrate 
would rather experience local spin-orbit fields
that can be very different for certain regions due to the different stacking configurations. 
Therefore, in Fig. \ref{Fig:SRT_grp_hBN}(a) we show a colormap of the 
calculated anisotropy $\xi$ as a function of $N$ and the energy 
when using the averaged parameters of graphene on hBN given in Tab. \ref{tab:fit_grp_hBN}. 
This averaged situation should correspond 
to a more realistic situation in a real heterostructure, 
where all kinds of stacking configurations are present simultaneously. 
We find that electrons have an anisotropy ratio $\xi \approx \frac{1}{2}$ 
almost independent of $N$ and the energy, see Fig. \ref{Fig:SRT_grp_hBN}(b), clearly different
from the pure ($\alpha$B, $\beta$H) configuration, compare to Fig. \ref{Fig:SRT_CCoBH}.
Close to the band edges, i.e. the charge neutrality point, the anisotropy can reach very large values. 
For holes the anisotropy varies around $\xi \approx 1$ for moderate doping densities.

So far, anisotropies of $\xi \approx 1$ have been measured for graphene 
on hBN and SiO$_2$ \cite{Guimaraes2014:PRL, Raes2017:PRB, Raes2016:NC, Ringer2018:PRB, Tombros2008:PRL}, 
in agreement with our averaged parameter results. 
A first indication of large anisotropies was 
found in hBN encapsulated bilayer graphene heterostructures \cite{Leutenantsmeyer2018:PRL, Xu2018:PRL}.
There it was shown, that the anisotropy $\xi$ decreases with 
increasing carrier density, in line with our results for monolayer graphene. 
They also showed that the anisotropy, at fixed doping level, 
can be strongly enhanced by an applied electric field.

\begin{figure}[htb]
 \includegraphics[width=.99\columnwidth]{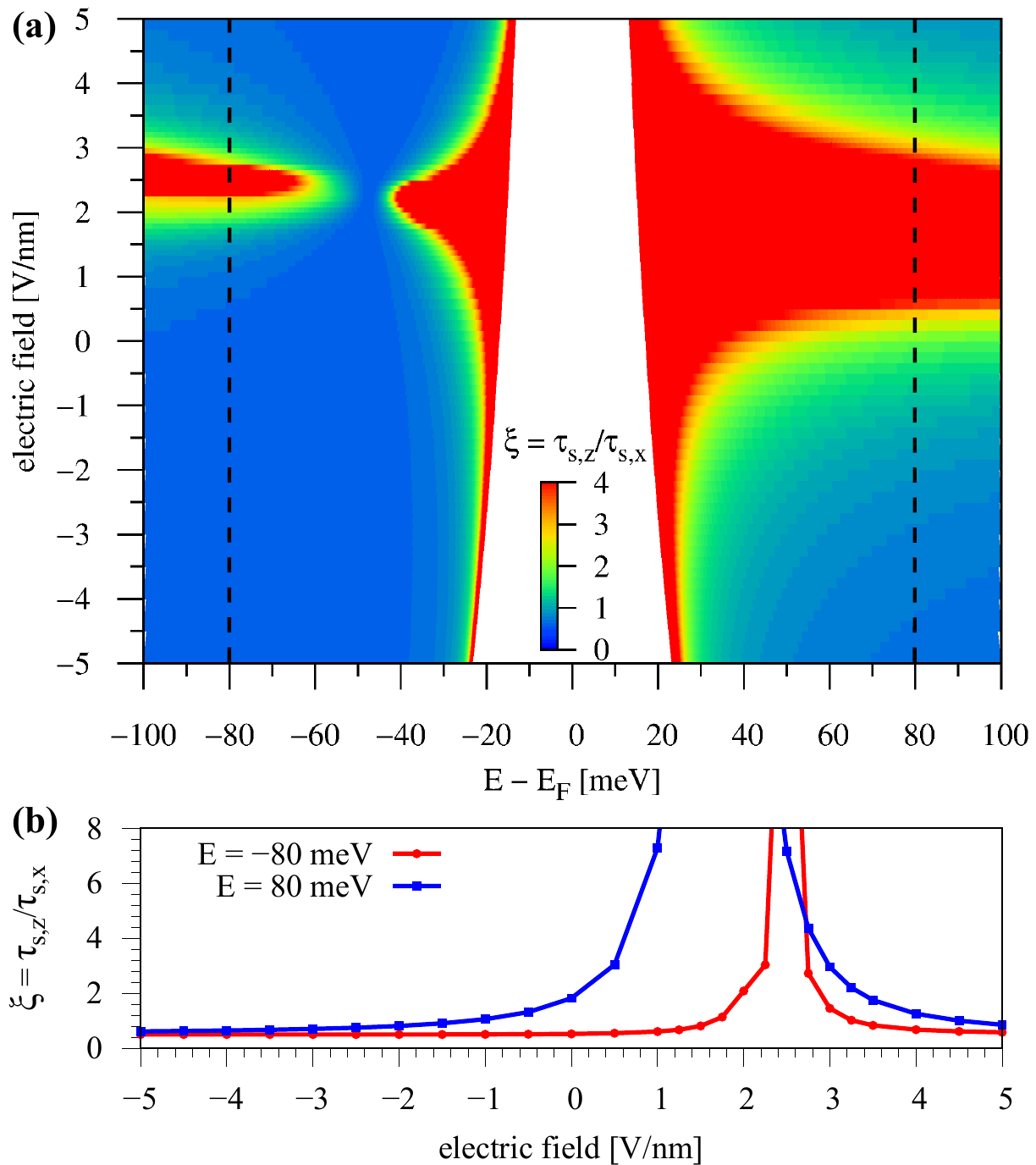}
 \caption{(Color online) (a) Calculated SR anisotropy $\xi = \tau_{s,\textrm{z}}/\tau_{s,\textrm{x}}$
  as a function of energy and applied transverse electric field for ($\alpha$B, $\beta$H) configuration, 
  using $\tau_p = 125$~fs and $\tau_{iv}= 8\cdot\tau_p$. 
  (b) Anisotropy $\xi$ at energies E $= \pm 80$~meV corresponding
  to the dashed lines in (a).
 }\label{Fig:anisotropy_grp_hBN_Efield}
\end{figure}

In dual gated structures, one can individually tune the doping level 
and the electric field across the heterostructure. 
In Fig. \ref{Fig:anisotropy_grp_hBN_Efield} we show the SR
anisotropy $\xi$, specifically for ($\alpha$B, $\beta$H) configuration
as a function of energy and applied transverse electric field, using the parameter sets 
for several finite electric field strengths, see Fig. \ref{Fig:Efield_CCoBH}.
We find that the anisotropy is strongly tunable by means of external gating. 
At around 2~V/nm we find a very strong enhancement
of the anisotropy, which is related to the zero transition of the Rashba SOC parameter. 
The anisotropy is giant for $\lambda_{\textrm{R}} \approx 0$, as the states are then mainly $s_z$ polarized. 
In Fig. \ref{Fig:anisotropy_grp_hBN_Efield}(b) we show that 
an electric field can tune the anisotropy 
by one order of magnitude at a fixed doping level.

\subsection{Additional considerations}
We now want to clarify two remaining issues: 
(i) Where does the discrepancy between the model and the first-principles data, see Fig. \ref{Fig:fitmodel_CCoBH}(b), come from? 
(ii) Is there a low-symmetry stacking configuration, where the orbital gap closes?
\subsubsection{Model Discrepancy}
\begin{figure}[htb]
 \includegraphics[width=.85\columnwidth]{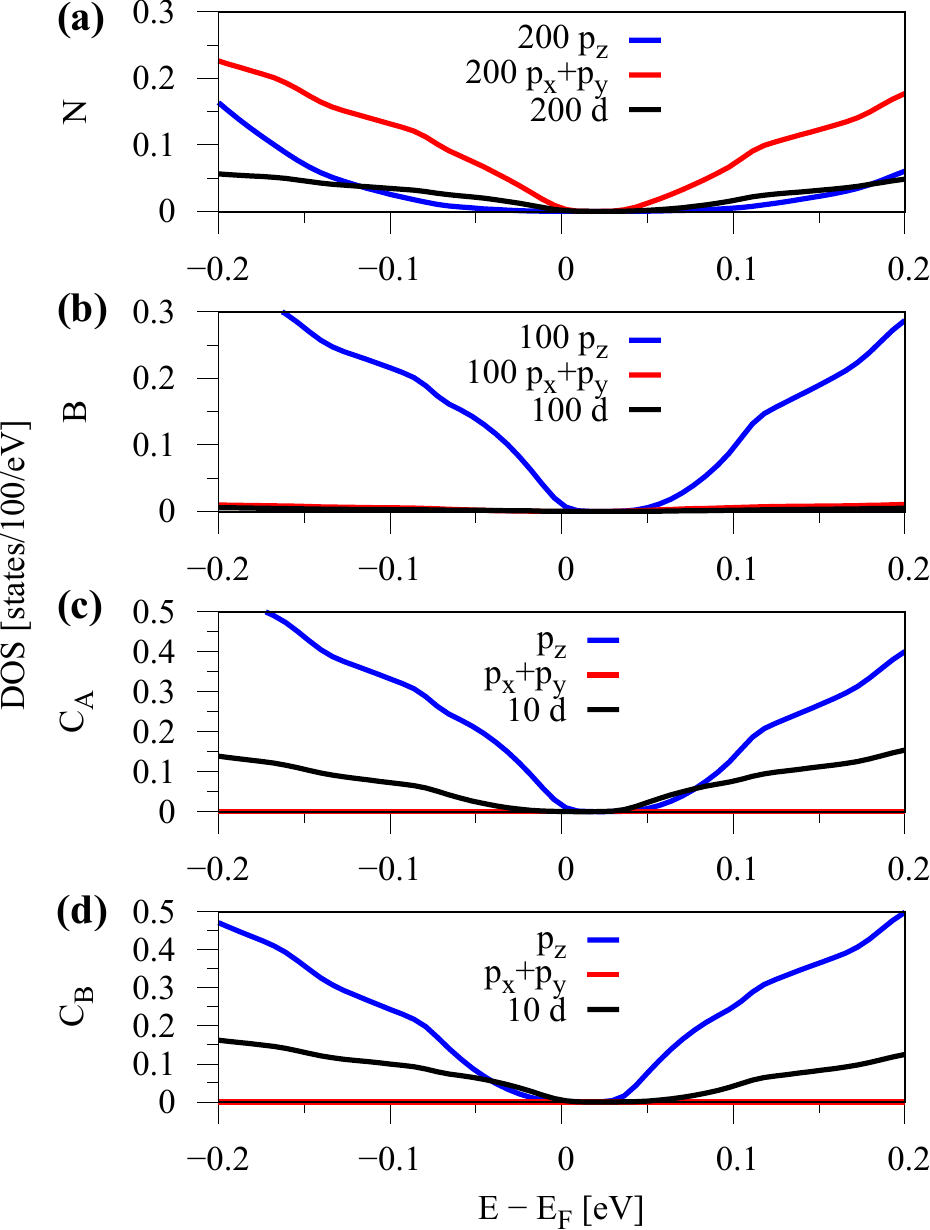}
 \caption{(Color online) Density of states of graphene on hBN around the Fermi level for ($\alpha$B, $\beta$H) configuration
and an interlayer distance of $3.35$~\AA. The DOS is multiplied by a factor of 100. Each subfigure (a)-(d) correspond to 
 a different atom. For each atom, the orbital contributions to the DOS are multiplied with the corresponding prefactor. 
 The DOS is calculated with a $k$-point grid of $180 \times 180 \times 1$. 
 }\label{Fig:FigDOS}
\end{figure}

In the case of the ($\alpha$B, $\beta$H) configuration, we have found that the splittings are overestimated (underestimated) along
the K-M (K-$\Gamma$) path, by the model.
The discrepancy in the splitting of the bands is due to the influence of the substrate. 
In general, the model Hamiltonian $\mathcal{H}$ just considers effective $\pi$ orbitals of graphene, however 
there seems to be a subtle influence from a hybridization to the $p$ orbitals of hBN.
If we look at the density of states (DOS) for the ($\alpha$B, $\beta$H) case, 
see Fig.~\ref{Fig:FigDOS}, we find that close to the Dirac point
there is a small contribution from nitrogen and boron $p$ states. 
Especially boron $p_z$ orbitals and nitrogen $p_x+p_y$ orbitals are contributing close to the charge neutrality point.
Moreover, from our distance study we find that the discrepancy between the model and the first-principles data
is getting smaller as we increase the interlayer distance. 

Finally, we calculate the low energy band structures, when SOC is artificially 
turned off on the nitrogen, boron, or carbon atoms, respectively.
The fit parameters for these situations are given in 
Tab.~\ref{tab:discrepancy}, along with the maximum discrepancy
for each situation. When SOC of the boron atom is turned off, 
the parameters and the fit accuracy are barely different.
A severe improvement of the fit is accomplished, 
when SOC of the nitrogen atom is turned off, reflected in 
the strongly reduced discrepancy between model and DFT data. 
Furthermore, if we turn off SOC on the carbon atoms of graphene, 
we can identify the contribution solely coming from the substrate, where we find 
negative intrinsic SOC parameters $\lambda_{\textrm{I}}^\textrm{A}$ and $\lambda_{\textrm{I}}^\textrm{B}$. 
Thus, nitrogen gives a non-negligible contribution to the SOC splitting of the Dirac bands. 
\begin{table*}[htb]
\begin{ruledtabular}
\begin{tabular}{l  c  c c  c  c   c  c  c c }
SOC on & $v_{\textrm{F}}/10^5 [\frac{\textrm{m}}{\textrm{s}}]$ & $\Delta$~[meV]& $\lambda_{\textrm{R}}~[\mu$eV] & $\lambda_{\textrm{I}}^\textrm{A}~[\mu$eV] &$\lambda_{\textrm{I}}^\textrm{B}~[\mu$eV] & $\lambda_{\textrm{PIA}}^\textrm{A}~[\mu$eV] & $\lambda_{\textrm{PIA}}^\textrm{B}~[\mu$eV] & discr. [a.u.]\\
\hline
N, B, C  & 8.308 & -17.08 & 10.65 & 5.00  & 9.37 & 33.58 & 37.57 & 1.265\\
N, C &  8.308 & -17.08 & 12.22 & 5.01 & 8.95 & 34.65 & 34.82 & 1.269 \\
B, C &  8.308 & -17.08 & 10.23 & 12.08 & 12.66 & -0.06 & -34.82 & 0.260\\
N, B &  8.308 & -17.07 & -1.82 & -7.09 & -2.79 & -9.53 & 66.55 & 1.349\\
C & 8.308 & -17.07 & 11.85 & 12.07 & 12.25 & -4.60 & -32.67 & 0.268\\
\end{tabular}
\end{ruledtabular}
\caption{\label{tab:discrepancy} Summary of the fitting parameters of Hamiltonian $\mathcal{H}$, 
for graphene on hBN for ($\alpha$B, $\beta$H) configuration and an interlayer distance of $3.35$~\AA. 
Here, we have artificially turned off SOC on nitrogen, boron, or carbon atoms, respectively.
 The Fermi velocity $v_{\textrm{F}}$, gap parameter $\Delta$, 
 Rashba SOC parameter $\lambda_{\textrm{R}}$, intrinsic SOC parameters $\lambda_{\textrm{I}}^\textrm{A}$ and $\lambda_{\textrm{I}}^\textrm{B}$ for sublattice A and B, and PIA SOC parameters 
 $\lambda_{\textrm{PIA}}^\textrm{A}$ and $\lambda_{\textrm{PIA}}^\textrm{B}$ for sublattice A and B. 
 The discrepancy is the calculated residual of the fit along the M-K-$\Gamma$ path given in arbitrary units.}
\end{table*}

From our analysis, we conclude that the discrepancy comes from 
nitrogen $p_x+p_y$ orbitals, that hybridize with $\pi$ orbitals of graphene.
Already such a very small contribution of $p_x+p_y$ orbitals, see Fig.~\ref{Fig:FigDOS}(a), 
can substantially influence the spin splitting and an effective model, based only on 
$\pi$ orbitals of graphene, can no longer perfectly describe the results. 
However, the overall fit is still very good and sufficient for our needs.

\subsubsection{Gap closing stacking}

\begin{figure}[htb]
 \includegraphics[width=.99\columnwidth]{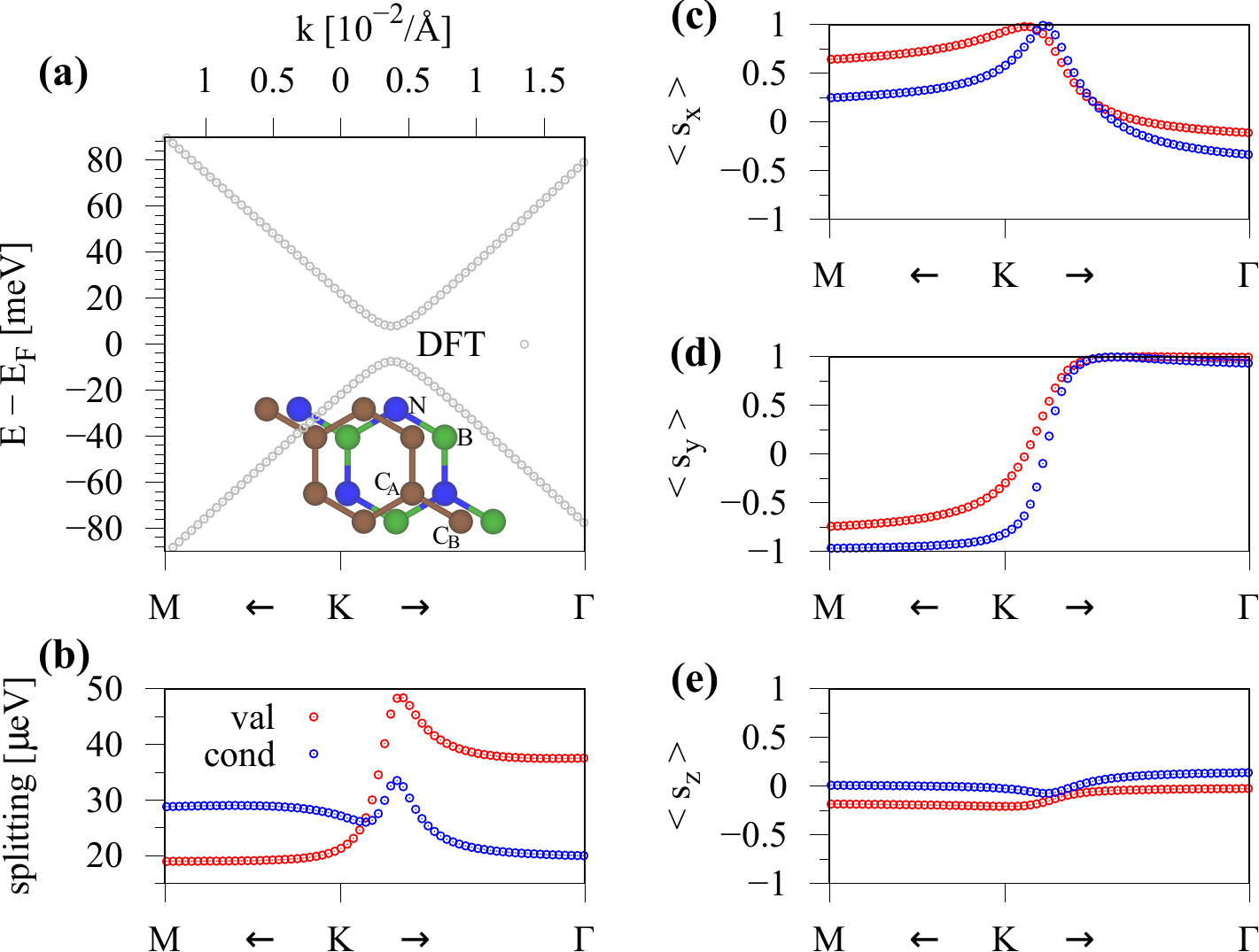}
 \caption{(Color online) Calculated band properties of graphene on hBN in 
 the vicinity of the K point and an interlayer distance of $3.45$~\AA. 
 (a) First-principles band structure and local stacking geometry. 
 (b) The splitting of conduction band $\Delta\textrm{E}_{\textrm{CB}}$ (blue) and valence band 
 $\Delta\textrm{E}_{\textrm{VB}}$ (red) close to the K point. 
 (c)-(e) The spin expectation values of the bands $\varepsilon_{2}^{\textrm{VB}}$ and $\varepsilon_{1}^{\textrm{CB}}$.
 }\label{Fig:bands_grp_hBN_gap_closing}
\end{figure}

We have seen that different stackings can lead to a different 
sign of the gap parameter $\Delta$, see Tab. \ref{tab:fit_grp_hBN}.
Consequently, as already mentioned, a local stacking geometry can exist, 
in a real moir\'{e} pattern geometry, that has a closed orbital gap.
In Fig. \ref{Fig:bands_grp_hBN_gap_closing} we show 
the low energy band properties 
of an arbitrary stacking geometry, without having any 
symmetry \footnote{The stacking is chosen such 
that each graphene sublattice has the same distance 
to an underlying boron and nitrogen atom. 
This seemed to be a good candidate for a gap closing, as 
the difference in the sublattice potential could vanish.}.

\begin{figure}[htb]
 \includegraphics[width=.99\columnwidth]{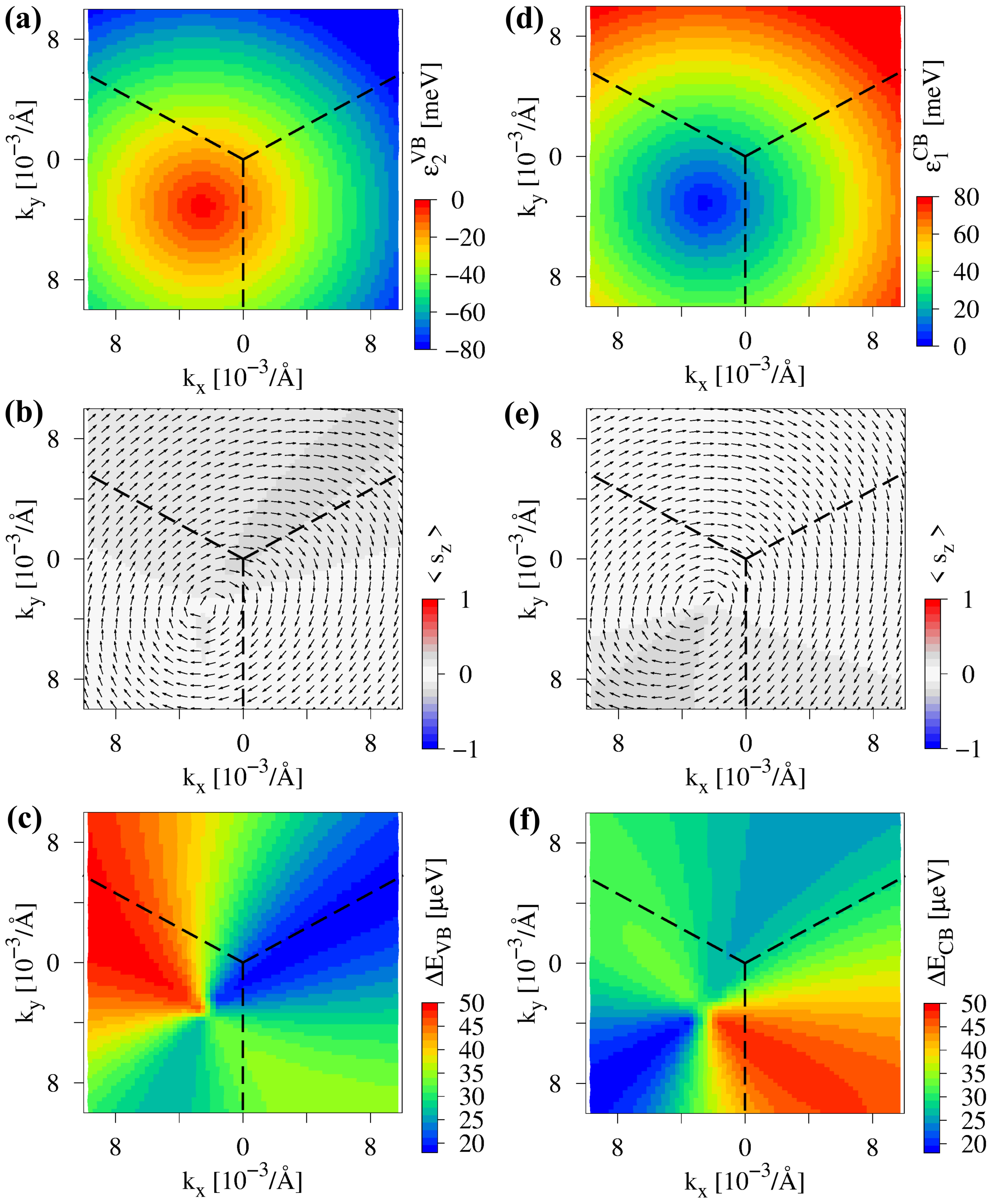}
 \caption{(Color online) Calculated low energy dispersion of graphene on hBN around the
 K point for stacking configuration in Fig. \ref{Fig:bands_grp_hBN_gap_closing} and an interlayer distance of $3.45$~\AA. 
 (a) 2D map of the energy of the valence band $\varepsilon_{2}^{\textrm{VB}}$, with the corresponding
 spin texture of the band shown in (b) and the splitting of the valence band  
 $\Delta\textrm{E}_{\textrm{VB}} = \varepsilon_{2}^{\textrm{VB}}-\varepsilon_{1}^{\textrm{VB}}$ shown in (c). 
 (d)-(f) The same as (a)-(c), but for conduction band $\varepsilon_{1}^{\textrm{CB}}$ and conduction band splitting
 $\Delta\textrm{E}_{\textrm{CB}} = \varepsilon_{2}^{\textrm{CB}}-\varepsilon_{1}^{\textrm{CB}}$. The dashed lines
 show the edges of the Brillouin zone with the K point at the center. 
 }\label{Fig:spinmap_grp_hBN_gap_closing}
\end{figure}

First of all, we notice that the Dirac point is no longer located at the K point. 
From the corresponding geometry in Fig. \ref{Fig:bands_grp_hBN_gap_closing}(a), 
we find that the hoppings, from say $\textrm{C}_{\textrm{A}}$ to the three 
nearest neighbors $\textrm{C}_{\textrm{B}}$, are all different 
due to the substrate. This asymmetry in the nearest neighbor 
hopping amplitudes leads to the shift of the Dirac 
point in momentum space \cite{Pereira2009:PRB, Wunsch2008:NJP}.
Since our model Hamiltonian considers only high-symmetry 
stacking configurations, without shifted Dirac cone, 
we cannot fit the data with it. 
From the spin expectation values we find a very pronounced 
Rashba spin-orbit field, as the $s_z$ component is strongly suppressed.

In order to identify the location of the Dirac point in momentum space, 
we calculate the dispersion as a 2D map in $k_x$-$k_y$ plane 
in the vicinity of the K point, see Fig. \ref{Fig:spinmap_grp_hBN_gap_closing}.
Indeed, we find that the Dirac point is shifted away from the corner of the Brillouin zone. 
At the Dirac point, the orbital gap is $1.64$~meV large. 
Due to the limited number of $k$ points in the calculation grid for the 2D map, 
we cannot identify the exact position of the Dirac point, so the orbital gap is not fully closed, 
but much smaller than in the high-symmetry stacking cases, see Tab. \ref{tab:fit_grp_hBN}. 
We also notice that the spin-orbit field is almost purely 
in-plane without any $s_z$ component, see Figs. 
\ref{Fig:spinmap_grp_hBN_gap_closing}(b) and \ref{Fig:spinmap_grp_hBN_gap_closing}(e), 
in a very large area around the Dirac point. 
Consequently, Rashba SOC plays an important role in this 
low-symmetry stacking configuration.  
If we look at the spin-orbit splitting 
of the bands, Figs. \ref{Fig:spinmap_grp_hBN_gap_closing}(c) and \ref{Fig:spinmap_grp_hBN_gap_closing}(f),
we find that there is no trigonal symmetry remaining. 
Such a stacking configuration completely breaks the symmetry of the graphene, 
due to the different hopping amplitudes between nearest neighbors caused by the hBN substrate. 
Of course, in a moir\'{e} geometry, several other stackings are present, that 
lead to very different local orbital gaps, spin-orbit fields, and spin splittings.

\section{hBN encapsulated graphene}
In this section we discuss the hBN/graphene/hBN heterostructures. 
We show our fit results to the low energy Hamiltonian 
for the different stacking configurations. 
Compared to the previous section, symmetry plays an important role
when fitting the Hamiltonian.
Again, we show the tunability of the parameters by applying a 
transverse electric field across the heterostructures. 
Finally we calculate SR times and anisotropies, and highlight 
differences to experimental findings in bilayer graphene.
\subsection{Low energy bands}

\begin{figure}[htb]
 \includegraphics[width=.98\columnwidth]{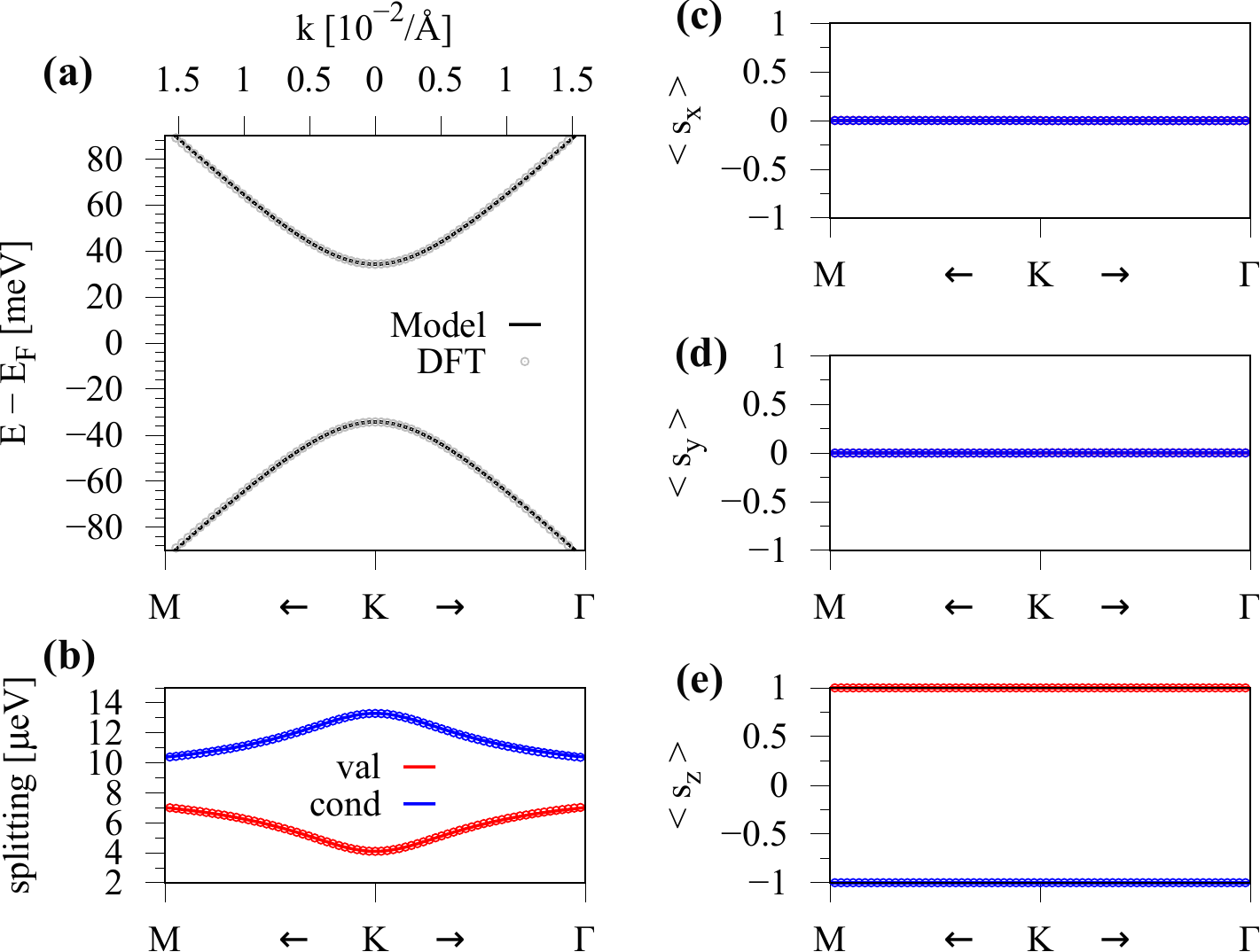}
 \caption{(Color online) Calculated band properties of hBN 
 encapsulated graphene in the vicinity of the K point 
 for (H$\alpha$H, B$\beta$B) = C1 configuration and interlayer 
 distances of $3.35$~\AA~between graphene and the hBN layers. 
 (a) First-principles band structure (symbols) with a fit to the model Hamiltonian (solid line). 
 (b) The splitting of conduction band $\Delta\textrm{E}_{\textrm{CB}}$ (blue) and valence band 
 $\Delta\textrm{E}_{\textrm{VB}}$ (red) close to the K point and calculated model results. 
 (c)-(e) The spin expectation values of the bands $\varepsilon_{2}^{\textrm{VB}}$ and 
 $\varepsilon_{1}^{\textrm{CB}}$ and comparison to the model results. 
 The fit parameters are given in Tab. \ref{tab:fit_hBN_grp_hBN}.
 }\label{Fig:fitmodel_HBoCCoHB}
\end{figure}
\begin{figure}[!htb]
 \includegraphics[width=.98\columnwidth]{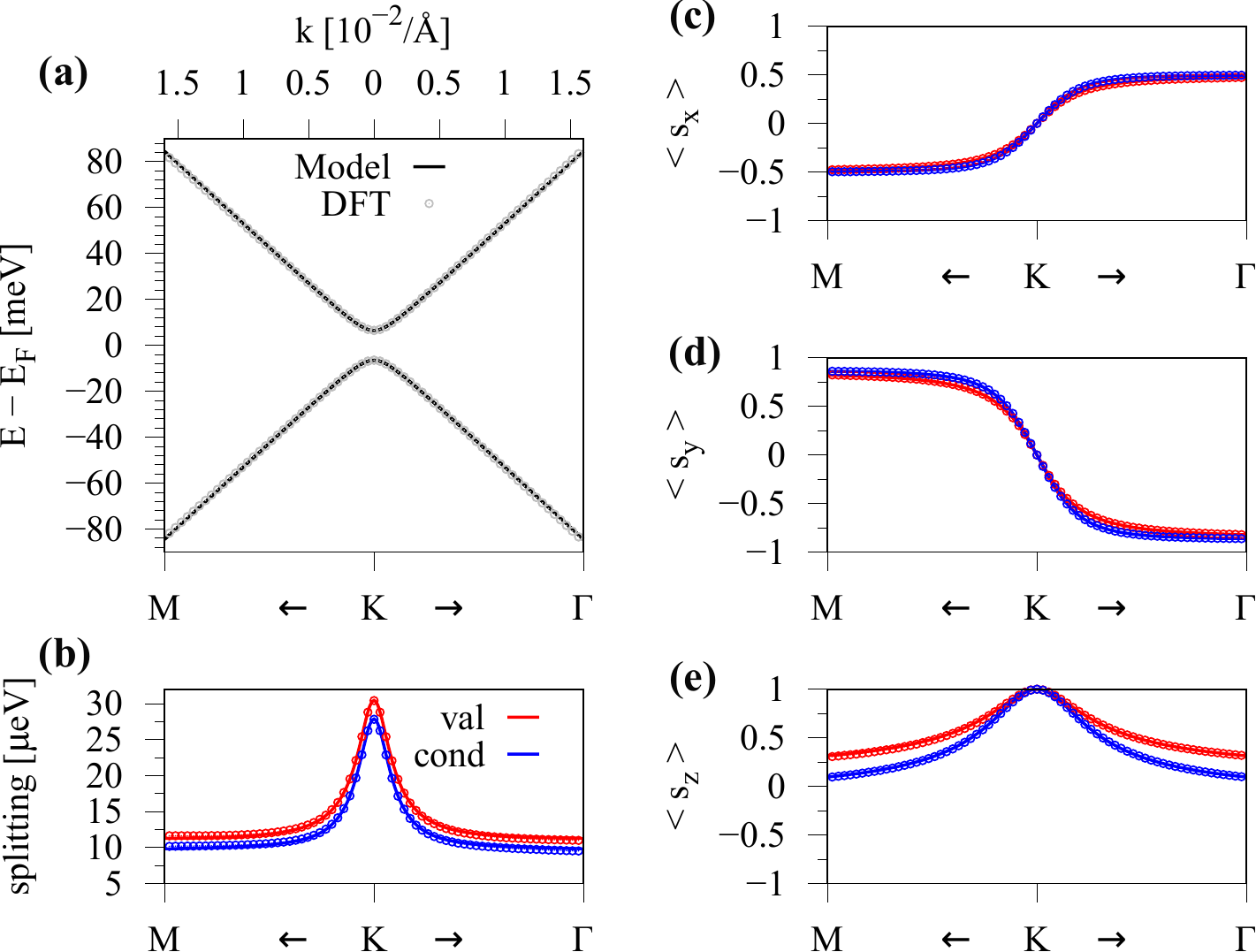}
 \caption{(Color online) Calculated band properties of hBN encapsulated graphene in the vicinity of the K point 
 for (B$\alpha$N, N$\beta$H) configuration with a distance of $3.55$~\AA~($3.50$~\AA) between graphene and the top (bottom) hBN layer. 
 (a) First-principles band structure (symbols) with a fit to the model Hamiltonian (solid line). 
 (b) The splitting of conduction band $\Delta\textrm{E}_{\textrm{CB}}$ (blue) and valence band 
 $\Delta\textrm{E}_{\textrm{VB}}$ (red) close to the K point and calculated model results. 
 (c)-(e) The spin expectation values of the bands $\varepsilon_{2}^{\textrm{VB}}$ and 
 $\varepsilon_{1}^{\textrm{CB}}$ and comparison to the model results. 
 The fit parameters are given in Tab. \ref{tab:fit_hBN_grp_hBN}.
 }\label{Fig:fitmodel_BNoCCoNH}
\end{figure}
\begin{figure}[!htb]
 \includegraphics[width=.98\columnwidth]{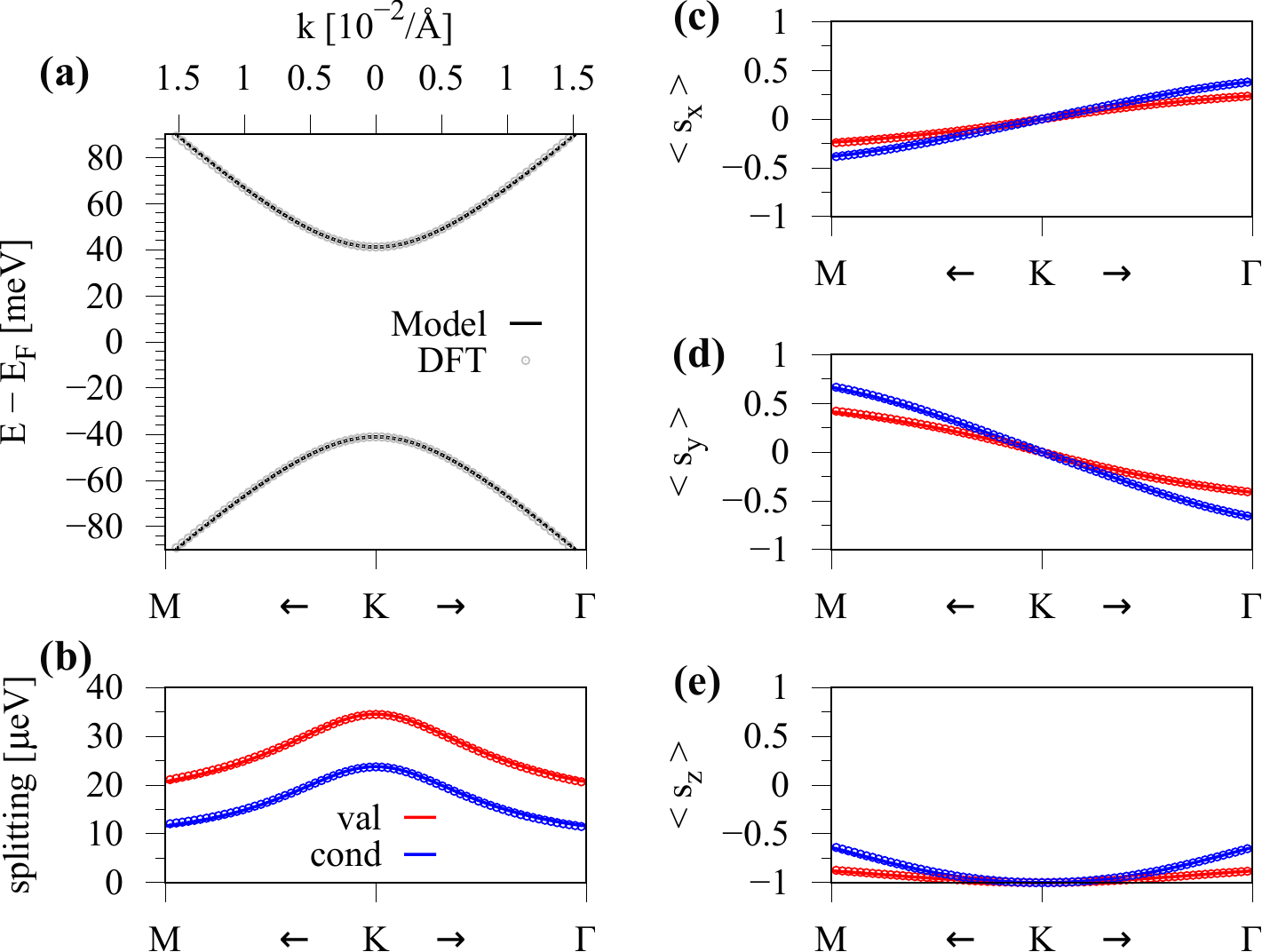}
 \caption{(Color online) Calculated band properties of hBN encapsulated graphene in the vicinity of the K point 
 for (N$\alpha$N, B$\beta$H) configuration with a distance of $3.55$~\AA~($3.50$~\AA) between graphene and the top (bottom) hBN layer. 
 (a) First-principles band structure (symbols) with a fit to the model Hamiltonian (solid line). 
 (b) The splitting of conduction band $\Delta\textrm{E}_{\textrm{CB}}$ (blue) and valence band 
 $\Delta\textrm{E}_{\textrm{VB}}$ (red) close to the K point and calculated model results. 
 (c)-(e) The spin expectation values of the bands $\varepsilon_{2}^{\textrm{VB}}$ and 
 $\varepsilon_{1}^{\textrm{CB}}$ and comparison to the model results. 
 The fit parameters are given in Tab. \ref{tab:fit_hBN_grp_hBN}.
 }\label{Fig:fitmodel_NBoCCoNH}
\end{figure}

\begin{table*}[htb]
\begin{ruledtabular}
\begin{tabular}{l  c  c c  c  c   c  c c}
Configuration & $\Delta E$ [meV]& $v_{\textrm{F}}/10^5~[\frac{\textrm{m}}{\textrm{s}}]$ & $\Delta$~[meV]& 
$\lambda_{\textrm{R}}~[\mu$eV] & $\lambda_{\textrm{I}}^\textrm{A}~[\mu$eV] &
$\lambda_{\textrm{I}}^\textrm{B}~[\mu$eV] & $\lambda_{\textrm{PIA}}^\textrm{A}~[\mu$eV] & 
$\lambda_{\textrm{PIA}}^\textrm{B}~[\mu$eV] \\
\hline
(H$\alpha$B, B$\beta$H) & 0.01 & 8.296 & 0&0  &2.19 &2.19 &0  &0 \\
(H$\alpha$N, N$\beta$H) & 26.60 & 8.068 & 0&0  &13.31 & 13.31 & 0 & 0\\
(N$\alpha$B, B$\beta$N) & 32.05 & 7.931 & 0 & 0 & 15.76 & 15.76 & 0 & 0 \\

(H$\alpha$H, B$\beta$B) = C1 & 0 & 8.294 & 34.24 & 0 & 6.65 & -2.05 & 0 & 0 \\
(N$\alpha$N, H$\beta$H) & 26.09 & 8.070 & 34.10 & 0 & 11.38 & 15.47 & 0 & 0 \\
(B$\alpha$B, N$\beta$N) & 31.76 & 7.932 & -48.00 & 0 & 18.95 & 12.34& 0 & 0\\

(H$\alpha$B, N$\beta$H) & 13.12 & 8.175 & -34.85 & -1.97 & 6.51& 9.05  & 3.15 & 31.35 \\
(N$\alpha$B, B$\beta$H) & 15.89 & 8.110 & 6.29 & -7.75 & 5.09 & 12.84 & 1.26  &22.72\\
(B$\alpha$N, N$\beta$H) & 29.20 & 7.998 & -6.50 & -4.97 & 15.23 & 13.92 & -61.34 & 49.22 \\

(N$\alpha$B, H$\beta$H) & 13.16 & 8.176 & -0.069 & -2.58 & 4.76 & 11.01 & 16.02 & 15.98\\
(B$\alpha$B, N$\beta$H) & 15.85 & 8.108 & -41.50& -7.37 & 8.31 & 9.55 & 6.71 & 16.36\\
(N$\alpha$N, B$\beta$H) & 28.82 & 8.000 & 41.14 &  -3.29 & 11.89 & 17.26 & 94.83 & -106.79 \\
\\
(H$\alpha$BN, B$\beta$HH) & 0.07 & 8.296 & 0.093 & 0.40 & 2.06 & 2.22 & 0 & 0   \\
(B$\alpha$BN, H$\beta$HH) & 0 & 8.298 & -34.06 & 0.31 & -2.23 & 6.66 & 0 & 0 \\

\end{tabular}
\end{ruledtabular}
\caption{\label{tab:fit_hBN_grp_hBN} Fit parameters for different hBN encapsulated graphene geometries, 
using the energetically most favorable graphene-hBN interlayer distances. 
The energy difference $\Delta E$ with respect to the C1 configuration,
 the Fermi velocity $v_{\textrm{F}}$, gap parameter $\Delta$, 
 Rashba SOC parameter $\lambda_{\textrm{R}}$,
 intrinsic SOC parameters $\lambda_{\textrm{I}}^\textrm{A}$ and $\lambda_{\textrm{I}}^\textrm{B}$, 
 and PIA SOC parameters  $\lambda_{\textrm{PIA}}^\textrm{A}$ and $\lambda_{\textrm{PIA}}^\textrm{B}$. 
 In the case of hBN/graphene/2hBN, the energy difference is with 
 respect to the (B$\alpha$BN, H$\beta$HH) configuration.}
\end{table*}

From our previous study of graphene on hBN, we already know what is the energetically most favorable distance
for each stacking geometry, which we keep for the encapsulated cases, respectively. 
Depending on the stacking of the top and bottom hBN with respect to the graphene, 
different interlayer distances can be present. 
The stacking sequences are defined in analogy to the graphene on hBN cases.
The energetically most favorable configuration is (H$\alpha$H, B$\beta$B), which we name C1 configuration.
According to this, we define several other configurations. 

For such a configuration, like (H$\alpha$H, B$\beta$B) = C1, 
we recover the mirror symmetry of graphene, see Fig.\ref{Fig:stackings}, 
reflected in the $D_{3h}$ symmetric version of the Hamiltonian with vanishing Rashba and PIA
contributions \cite{Kochan2017:PRB}.
In Fig. \ref{Fig:fitmodel_HBoCCoHB}, we show the low energy band properties 
of the C1 configuration, along with a fit to our model Hamiltonian. 
We can see perfect agreement with the first-principles data, just using the four parameters $v_{\textrm{F}}$,
$\Delta$, $\lambda_{\textrm{I}}^\textrm{A}$, and $\lambda_{\textrm{I}}^\textrm{B}$. 
Rashba and PIA SOC parameters are not necessary and strictly zero for the fit, 
especially for this mirror symmetric configuration, as explained.
Therefore the bands are purely $s_z$ polarized. 

In Figs. \ref{Fig:fitmodel_BNoCCoNH} and \ref{Fig:fitmodel_NBoCCoNH}, we show the low energy band properties 
of the (B$\alpha$N, N$\beta$H) and (N$\alpha$N, B$\beta$H) configurations, along with a fit to our 
model Hamiltonian, as further examples of the robustness of the Hamiltonian. 
We can see again perfect agreement with the first-principles data. 
Even though the low energy band properties are somewhat similar, each configuration has a very individual
parameter set. 

The parameters, best fitting the DFT results,
are given in Tab. \ref{tab:fit_hBN_grp_hBN}.
We find that there is another configuration, (H$\alpha$B, B$\beta$H), having almost
the same total energy as the (H$\alpha$H, B$\beta$B) one, in agreement with literature \cite{Quhe2012:NPG}. 
Overall, the magnitudes of SOCs are tens of $\mu$eV, while the parameters can differ 
(also in sign) from structure to structure.
For example, the (H$\alpha$B, B$\beta$H) configuration is in the $D_{3d}$ subgroup and 
the only allowed SOC parameters are 
$\lambda_{\textrm{I}}^\textrm{A} = \lambda_{\textrm{I}}^\textrm{B} = \lambda_{\textrm{I}}$. 
In this case the orbital gap $\Delta = 0$, since the overall potential, 
from top and bottom hBN layer, is equal for the two graphene sublattices. 
When $\Delta = 0$, the spectrum opens a gap due to SOC with degenerate CB and VB, 
just as for pristine graphene \cite{Gmitra2009:PRB}. 
Also worth noticing is, that if one takes the average of 
$\lambda_{\textrm{I}}^\textrm{A}$ and $\lambda_{\textrm{I}}^\textrm{B}$ of the C1 configuration,
you arrive at the parameter $\lambda_{\textrm{I}}$ for the (H$\alpha$B, B$\beta$H) configuration. 
In addition, some encapsulated configurations show a negative Rashba SOC parameter. 
This means that the spin-orbit field rotates in a counterclockwise direction, 
in contrast to the graphene/hBN cases, see for example Fig. \ref{Fig:spinmap_CCoBH}.

In real systems, one expects that all of these configurations are present at the same time, 
due to the moir\'{e} pattern that is formed as a consequence of 
slightly different lattice constants of graphene and hBN. 
In addition stacking configurations can occur that, locally, have no symmetry at all, 
as shown in the previous section, making these heterostructures 
quite complicated to describe on the global scale. 
In experiments, also asymmetric hBN encapsulated graphene structures are used, 
say with two hBN layers below graphene and one hBN layer above it. 
We also calculate this scenario, especially for the configurations that are energetically most favorable. 
The stacking of hBN itself is (BN, NB) (boron over nitrogen, nitrogen over boron) and we 
take a distance of $3.35$~\AA~between the hBN layers. 
These configurations and their fit parameters are also summarized in Tab. \ref{tab:fit_hBN_grp_hBN},
with the naming convention in analogy to the other cases. 
Comparing the results of this asymmetric encapsulations with the corresponding symmetric encapsulations --- 
for example compare (H$\alpha$H, B$\beta$B) and (B$\alpha$BN, H$\beta$HH) --- 
we find that they are almost the same. The only thing is that
the C$_\textrm{A}$ and C$_\textrm{B}$ sublattice are interchanged in these two configurations, which is 
reflected in the intrinsic SOC parameters. 

One of the main conclusions is, that in the case of hBN encapsulated graphene, 
the average Rashba SOC parameter is 
reduced in contrast to the graphene/hBN average Rashba parameter, 
while intrinsic SOC parameters have similar magnitudes. 
Therefore, in spin transport, hBN encapsulated graphene should have
longer SR times for out-of-plane spins.

\subsection{Transverse Electric Field}

\begin{figure}[htb]
 \includegraphics[width=.99\columnwidth]{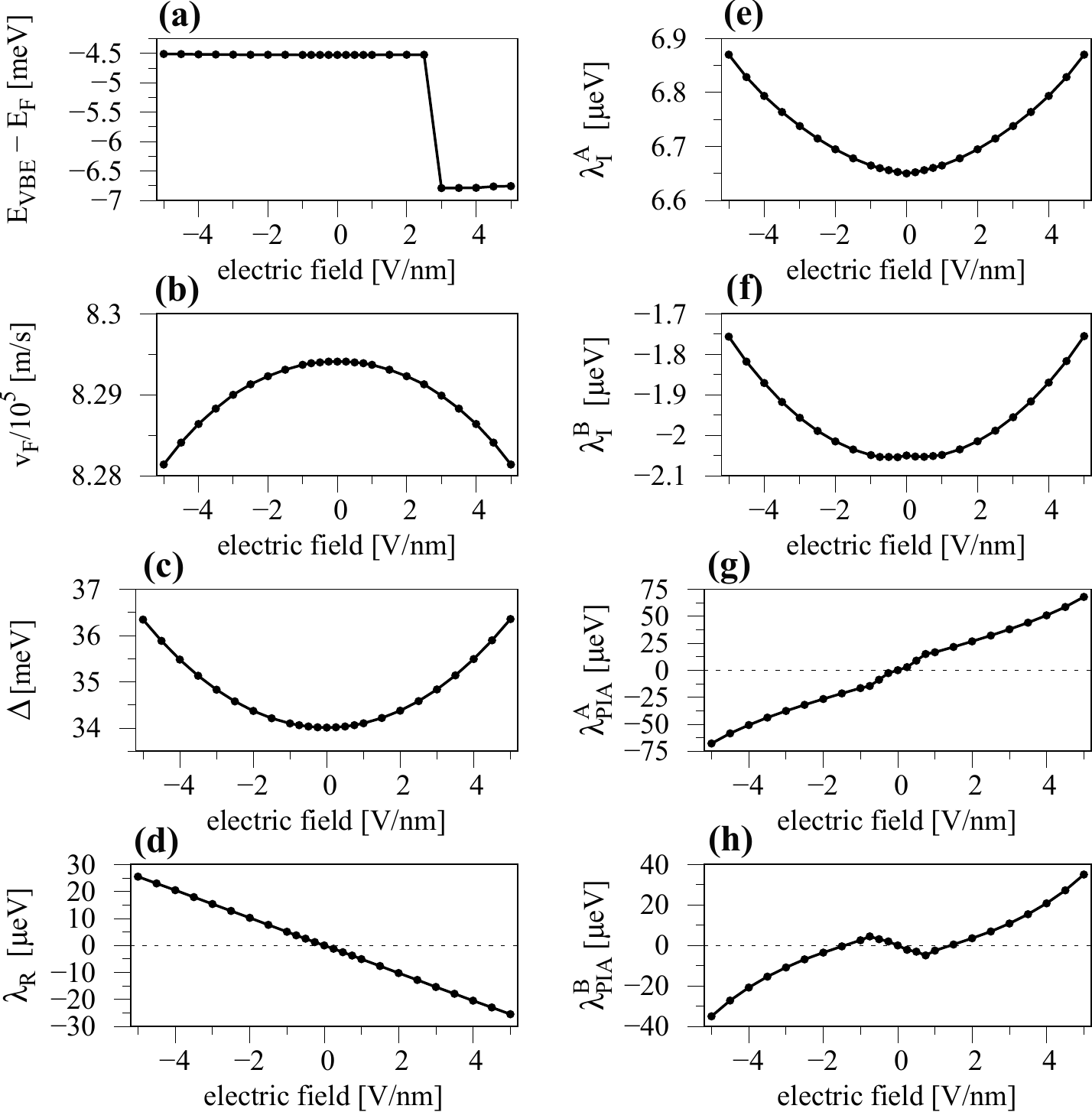}
 \caption{(Color online) Fit parameters as a function of the 
 applied transverse electric field for the C1 configuration. 
 (a) Valence band edge with respect to the Fermi level, 
 (b) Fermi velocity $v_{\textrm{F}}$, (c) gap parameter $\Delta$, 
 (d) Rashba SOC parameter $\lambda_{\textrm{R}}$, 
 (e,f) intrinsic SOC parameters $\lambda_{\textrm{I}}^\textrm{A}$
  and $\lambda_{\textrm{I}}^\textrm{B}$, and (g,h) PIA SOC parameters 
 $\lambda_{\textrm{PIA}}^\textrm{A}$ and $\lambda_{\textrm{PIA}}^\textrm{B}$.
 }\label{Fig:Efield_HBoCCoHB}
\end{figure}

In the case of hBN encapsulated graphene we consider the C1 configuration. 
The tunability of the parameters for C1, by electric field, is shown in Fig. \ref{Fig:Efield_HBoCCoHB}.
The Rashba and PIA parameters --- which are due to inversion asymmetry --- 
are odd functions (almost linear) of electric field and strongly tunable.
In contrast, the orbital parameters $v_{\textrm{F}}$ and $\Delta$, 
as well as intrinsic SOC parameters $\lambda_{\textrm{I}}^\textrm{A}$ and $\lambda_{\textrm{I}}^\textrm{B}$,
are even functions and only weakly affected by the field. 
Applying an external field, we find a linear tunability of roughly 5~$\mu$eV per V/nm of $\lambda_{\textrm{R}}$, 
similar to freestanding graphene \cite{Gmitra2009:PRB}, which is expected for the mirror-symmetric C1.

\begin{figure}[htb]
 \includegraphics[width=.98\columnwidth]{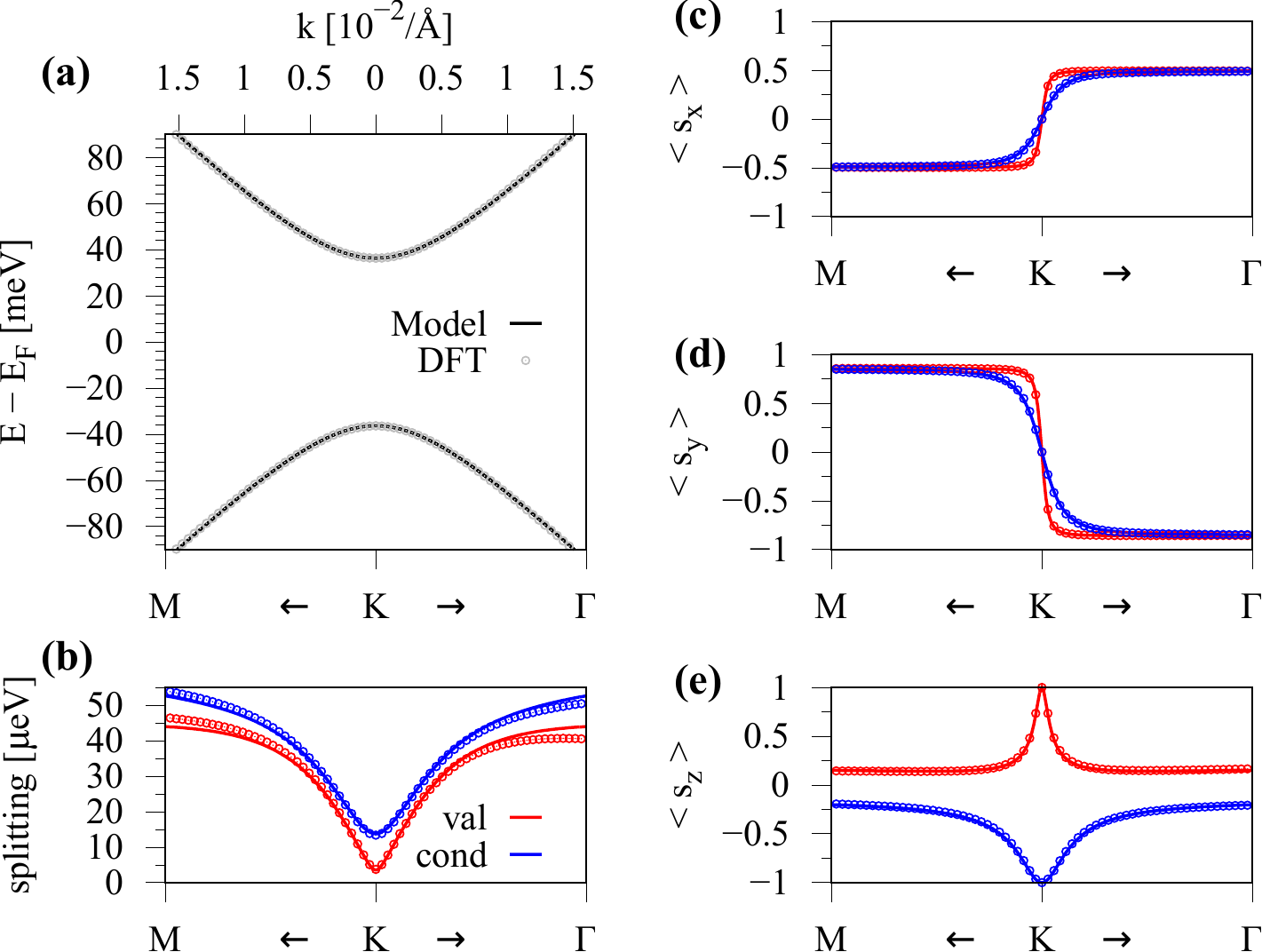}
 \caption{(Color online) Calculated band properties of hBN encapsulated graphene in the vicinity of the K point 
 for C1 configuration and interlayer distances of $3.35$~\AA~between graphene and the hBN layers
 with external electric field of 5~V/nm. 
 (a) First-principles band structure (symbols) with a fit to the model Hamiltonian (solid line). 
 (b) The splitting of conduction band $\Delta\textrm{E}_{\textrm{CB}}$ (blue) and valence band 
 $\Delta\textrm{E}_{\textrm{VB}}$ (red) close to the K point and calculated model results. 
 (c)-(e) The spin expectation values of the bands $\varepsilon_{2}^{\textrm{VB}}$ and $\varepsilon_{1}^{\textrm{CB}}$ and comparison to the model results. The fit parameters are given in the main text.
 }\label{Fig:fitmodel_HBoCCoHB_E5Vnm}
\end{figure}

In Fig.~\ref{Fig:fitmodel_HBoCCoHB_E5Vnm} we show the low energy band properties of the 
C1 configuration, along with a fit to our model Hamiltonian with applied external electric field of 5 V/nm.
Comparing the two results, Figs. \ref{Fig:fitmodel_HBoCCoHB} and \ref{Fig:fitmodel_HBoCCoHB_E5Vnm},
we can clearly see that the orbital low energy band structure looks the very same. 
However, the band splittings away from the K point are strongly enhanced and the spin expectation values show a
clear signature of Rashba SOC. The application of a realistic electric field 
of 5 V/nm enhances the spin-orbit band splittings
by a factor of 5 away from the K point.
This has substantial influence on the spin lifetimes and SR anisotropies.

\subsection{Spin Relaxation Anisotropy}

\begin{figure}[!htb]
 \includegraphics[width=.99\columnwidth]{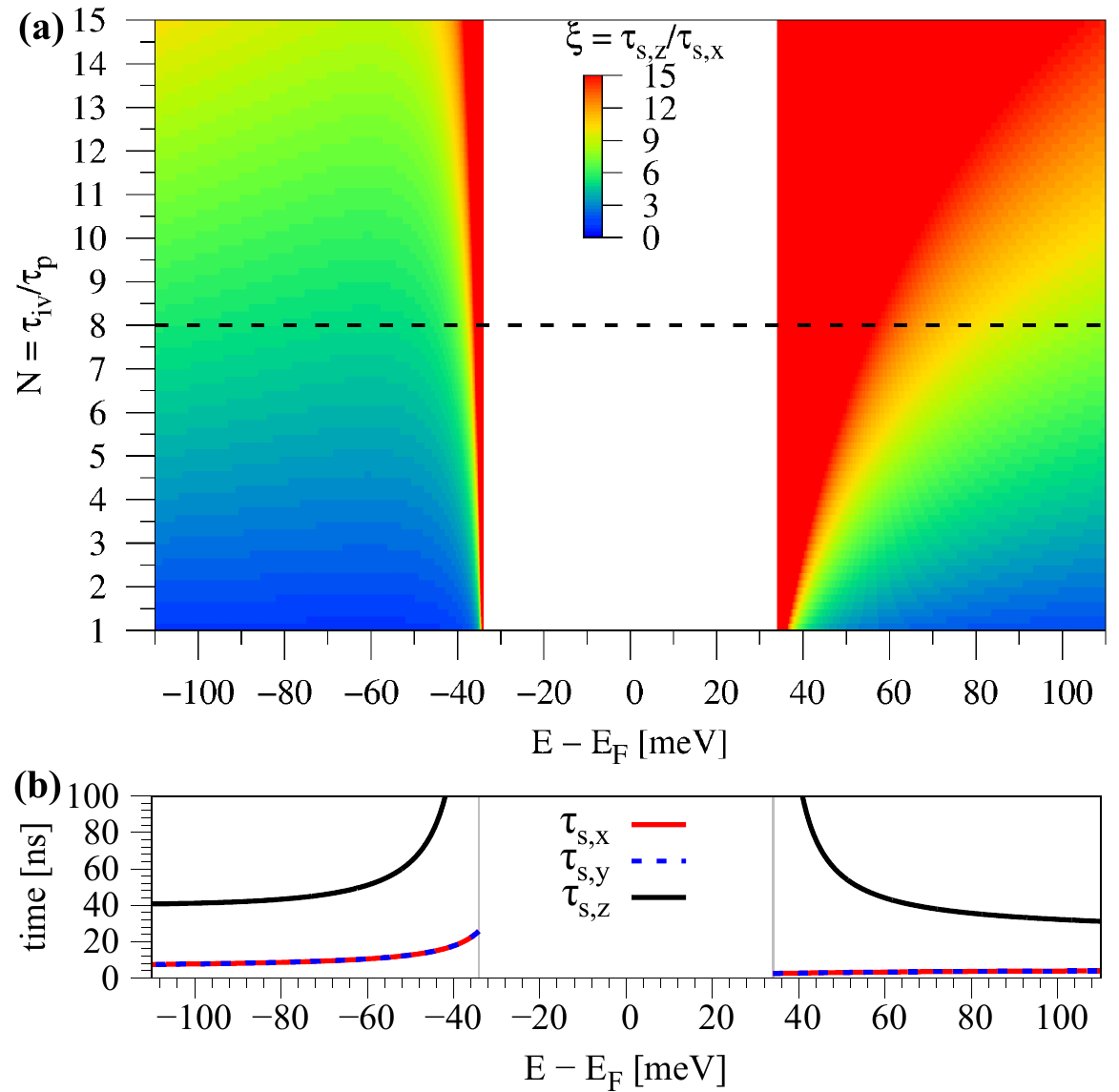}
 \caption{(Color online) Calculated SR times and anisotropies 
 for C1 configuration and an electric field of 1~V/nm. 
 (a) SR anisotropy $\xi = \tau_{s,\textrm{z}}/\tau_{s,\textrm{x}}$ 
 as a function of $N = \tau_{iv}/\tau_p$ and the energy. (b) Individual SR times as a function of energy 
corresponding to the dashed line in (a) with $\tau_p = 125$~fs and $\tau_{iv}= 8\cdot\tau_p$. 
The gray lines indicate the band edges. 
 }\label{Fig:SRT_HBoCCoHB}
\end{figure}

While experimental spectral sensitivities approach the limits of tens of $\mu$eV for encapsulated graphene, 
making the above calculations relevant for sensitive mesoscopic transport measurements, the most striking
ramifications of the obtained spin-orbit tunability is expected to be in SR anisotropy, which has
been a hotly debated issue recently. Indeed, we predict a wide electrical tunability of the SR time 
of this basic structure. Similar to the previous section, we calculate the SR times and anisotropies
for the selected lowest energy C1 configuration.

For completeness, we first show the SR anisotropy of C1, in Fig. \ref{Fig:SRT_HBoCCoHB}, as
a function of $N = \tau_{iv}/\tau_p$ and the energy for fixed electric field of 1~V/nm. 
The individual SR times are up to $100$~ns close to the band edges, see Fig. \ref{Fig:SRT_HBoCCoHB}(b), 
due to weak Rashba SOC in hBN encapsulated graphene.
The anisotropy is giant close to the band edges and decreases with increasing the doping.
Depending on the exact value of $\tau_{iv}$ and the doping,
the anisotropy can change within one order of magnitude. 

\begin{figure}[htb]
 \includegraphics[width=.99\columnwidth]{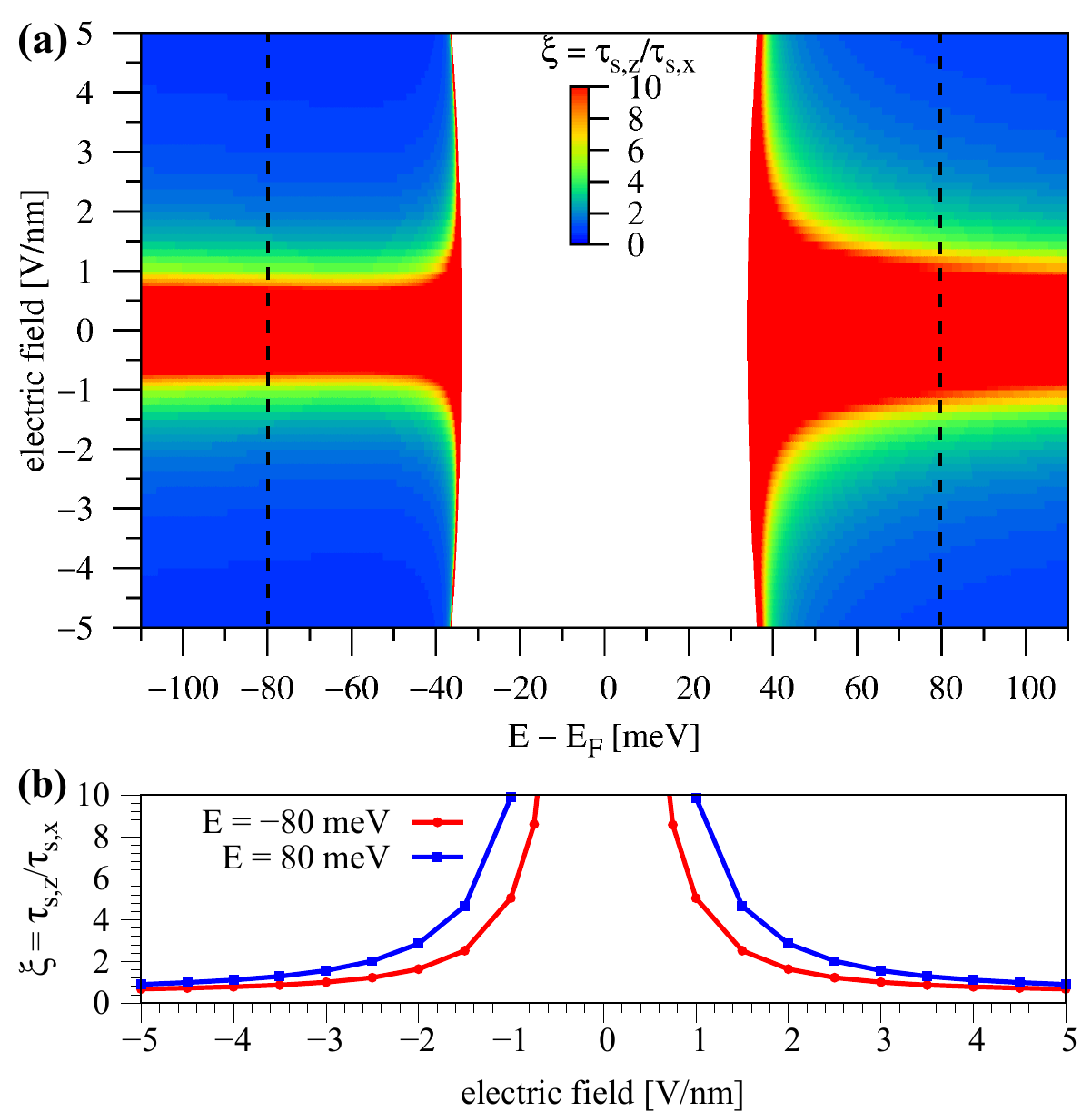}
 \caption{(Color online) (a) Calculated SR anisotropy $\xi = \tau_{s,\textrm{z}}/\tau_{s,\textrm{x}}$
  as a function of energy and applied transverse electric field for C1 configuration of hBN encapsulated graphene, 
  using $\tau_p = 125$~fs and $\tau_{iv}= 8\cdot\tau_p$. (b) Anisotropy $\xi$ at energies E $= \pm 80$~meV corresponding
  to the dashed lines in (a).   
 }\label{Fig:anisotropy_hBN_grp_hBN_Efield}
\end{figure}

In Fig. \ref{Fig:anisotropy_hBN_grp_hBN_Efield} we show the SR anisotropy $\xi$, 
specifically for C1, as a function of energy and transverse electric field.
We find that the anisotropy is strongly tunable by both the field and the doping level. 
At 0~V/nm the anisotropy is giant due to $\lambda_{\textrm{R}} = 0$, 
as shown in Fig. \ref{Fig:Efield_HBoCCoHB}(d), so that spin-orbit fields are out of plane. 
As the applied field increases, the anisotropy decreases. Overall, the anisotropy can be
tuned electrically from the usual 2D Rashba limit (0.5), to the opposite case of strong 
out-of-plane fields ($\xi \gg 1$), {\it for a fixed doping}, see 
Fig. \ref{Fig:anisotropy_hBN_grp_hBN_Efield}(b).
This is an unprecedented tunability for an electronic system. 

Curiously, the SR anisotropy decreases with increasing electric field (for a fixed doping), while
in encapsulated bilayer graphene the opposite was found \cite{Xu2018:PRL}. The reason is that 
in bilayer graphene the spin splitting at K is not tunable beyond a certain threshold \cite{Konschuh2012:PRB},
at marked contrast to monolayer graphene.

\section{Summary \& Conclusions}

In summary, we were able, by combining extensive first-principles calculations 
and a minimal tight-binding model, to extract useful orbital and spin-orbit coupling parameters for 
(hBN)/graphene/hBN heterostructures.
The extracted parameters depend on stacking configurations, interlayer distances, and a transverse electric field,
giving a rich playground for spin physics.
The consideration of different stacking configurations is important for realistic moir\'{e} pattern 
geometries of graphene and hBN. Spin-orbit fields in graphene, and consequently spin transport, 
can be controlled by the application of a transverse electric field.
Finally, the calculated SR times exhibit giant and tunable anisotropies, which 
are experimentally testable fingerprints of the ultimate role of SOC in SR
in graphene.

\acknowledgments
This work was supported by DFG SPP 1666, SFB 1277 (A09), 
the European Unions Horizon 2020 research and innovation program under Grant No. 785219, 
and by the reintegration scheme MSVVaS SR 90/CVTISR/2018 and VVGS-2018-887.

\bibliography{paper}

\end{document}